\documentclass[english,aps,pra,reprint]{revtex4-1}
\pdfoutput=1
\usepackage[T1]{fontenc}
\usepackage[latin9]{inputenc}
\usepackage{color}
\usepackage{babel}
\usepackage{bm}
\usepackage{amsmath}
\usepackage{amssymb}
\usepackage{graphicx}
\usepackage[unicode=true,pdfusetitle,
 bookmarks=true,bookmarksnumbered=false,bookmarksopen=false,
 breaklinks=false,pdfborder={0 0 0},pdfborderstyle={},backref=false,colorlinks=true]
 {hyperref}
\hypersetup{
 pdfborderstyle={},pdfborderstyle={},linkcolor=blue,citecolor=blue,urlcolor=blue}

\makeatletter

\usepackage{color}
\usepackage{babel}
\usepackage{bm}

\usepackage{babel}

\makeatother

\begin{document}
\global\long\def\pd{\partial}

\global\long\def\Tr{\textrm{Tr}}

\global\long\def\Id{\textrm{Id}}

\global\long\def\re{\textrm{Re}}

\global\long\def\im{\textrm{Im}}

\global\long\def\sgn{\textrm{sgn}}

\global\long\def\pv{\text{P}}

\global\long\def\Ang{\text{ Å}}

\global\long\def\dag{\dagger}

\global\long\def\uua{\uparrow}

\global\long\def\dda{\downarrow}

\global\long\def\ra{\rightarrow}

\global\long\def\lla{\leftarrow}

\global\long\def\x{\times}

\global\long\def\kp{\otimes}

\title{General theoretical description of angle-resolved photoemission spectroscopy
of van der Waals structures}

\author{B. Amorim}

\affiliation{CeFEMA, Instituto Superior Técnico, Universidade de Lisboa, Av. Rovisco
Pais, 1049-001 Lisboa, Portugal}
\email{bruno.a.c.amorim@tecnico.ulisboa.pt; amorim.bac@gmail.com}

\begin{abstract}
We develop a general theory to model the angle-resolved photoemission
spectroscopy (ARPES) of commensurate and incommensurate van der Waals
(vdW) structures, formed by lattice mismatched and/or misaligned stacked
layers of two-dimensional materials. The present theory is based on
a tight-binding description of the structure and the concept of generalized
umklapp processes, going beyond previous descriptions of ARPES in
incommensurate vdW structures, which are based on continuous, low-energy
models, being limited to structures with small lattice mismatch/misalignment.
As applications of the general formalism, we study the ARPES bands
and constant energy maps for two structures: twisted bilayer graphene
and twisted bilayer MoS$_{2}$. The present theory should be useful
in correctly interpreting experimental results of ARPES of vdW structures
and other systems displaying competition between different periodicities,
such as two-dimensional materials weakly coupled to a substrate and
materials with density wave phases.
\end{abstract}
\maketitle

\section{Introduction}

The development of fabrication techniques in recent years enabled
the creation of structures formed by stacked layers of different two-dimensional
(2D) materials, referred to as van der Waals (vdW) structures \citep{Ponomarenko_2011,Novoselov_2012,Geim_2013}.
By combining layers of materials displaying different properties,
it is possible to engineer devices with new functionalities, not displayed
by the individual layers. This makes vdW structures very appealing
from the applications point of view. As examples, transistors based
on graphene and hexagonal boron nitride ($h$-BN) or a semiconducting
transition metal dichalcogenide (STMD) \citep{Britnell_2012,Georgiou_2012}
and photodetectors based on graphene and a STMD \citep{Britnell_2013,Liu_2013,Massicotte_2015}
have already been realized. The properties of a vdW structure depend
not only on the properties of the individual layers, but also on how
different layers interact with each other. Due to the high crystallographic
quality of 2D materials, the interlayer interaction depends crucially
on the lattice mismatch and misalignment between different layers.
This is clearly exemplified, both experimentally \citep{Mishchenko_2014}
and theoretically \citep{Brey_2014,Amorim_2016}, by the observation
of negative differential conductance in graphene/$h$-BN/graphene
vertical tunneling transistors, where the bias voltage at which peak
current occurs is controlled by the angle between the misaligned graphene
electrodes. A necessary step to fully understand and take advantage
of vdW structures is to characterize their electronic properties and
how these depend on the lattice mismatch/misalignment.

Angle-resolved photoemission spectroscopy (ARPES) is an extensively
used tool to characterize the electronic degrees of freedom of materials
\citep{Damascelli_2003,Kordyuk_2014,Moser_2017}. In crystals, ARPES
is generally understood as a direct probe of the electronic band structure
over the Brillouin zone of occupied states. Nevertheless, even in
a perfect crystal where the notions of reciprocal space and Brillouin
zone are well defined, this picture might breakdown, as the ARPES
response is weighted by matrix elements which describe the light induced
electronic transition from a crystal bound state to a photoemitted
electron state. For bands that are well decoupled from the remaining
band structure, the ARPES matrix elements are featureless, and indeed
ARPES can be seen as a direct probe of the band structure. However,
exceptions to this can occur and the matrix elements can impose selection
rules on the transitions. Two well known examples where this occurs
are graphite \citep{Shirley_1995} and graphene \citep{Bostwick_2007,Bostwick_2007b,Mucha_2008,Gierz_2010,Jung_2010,Puschnig_2015,Lee_2017}.
In these materials part of the Fermi surface is not observed in constant
energy ARPES map \footnote{As shown in Ref.~\citep{Gierz_2011} it is actually possible to observe
the full Fermi surface of graphene in ARPES by using $s$-polarized
incident light. This phenomenon can be explained theoretically if
scattering of the photoemitted state by the graphene lattice is taken
into account. Such effects will not be considered in the present work. }. This effect is due to the ARPES matrix elements, which suppress
the signal from some parts of the band structure.

Another case where the ARPES matrix elements should play an important
role is in systems with competing periodicities, such as materials
displaying charge density wave (CDW) phases \citep{Voit_2000,Chen_2015}.
In this case it is easy to understand how the interpretation of ARPES
as a direct probe of the band structure can break down. Let us suppose
that for a given material in the normal, undistorted phase, ARPES
accurately maps the electronic band structure. Suppose that the system
undergoes a transition into a commensurate CDW, with a larger unit
cell. If the distortion is small, the bands in Brillouin zone will
be weakly perturbed apart from back-folding into the new, smaller,
Brillouin zone associated with the enlarged unit cell. If the CDW
perturbation is weak, by an adiabatic argument, the ARPES mapped bands
observed in both phases must be essentially unchanged. This means
that the signal of the back-folded bands must be very weak, and the
observed ARPES bands will mostly follow the bands of the undistorted
phase, in an extended zone scheme. The suppression of the back-folded
bands is encoded by the ARPES matrix elements. This was exemplified
in Ref.~\citep{Voit_2000} in a simple one-dimensional tight-binding
model. These effects may also be relevant when interpreting ARPES
experiments in cuprates, for which a hidden density wave state has
been proposed \citep{Chakravarty_2003}.

In vdW structures, such competing periodicities naturally occur due
to the lattice mismatch between different layers. Therefore, a general
theory capable of correctly taking into account ARPES matrix elements
is essential to interpret ARPES data from vdW structures. We point
out that the modeling of ARPES in twisted bilayer graphene \citep{Pal_2013}
and graphene/$h$-BN \citep{Mucha_2016} structures has been considered
previously in the literature. However the models employed relied on
effective low energy, continuous descriptions of the systems, which
are only valid for small misalignment angles. As the field of vdW
structures develops, a more general and flexible approach is required.
The goal of this work is to develop a general framework to theoretically
model ARPES, which is valid for both commensurate and incommensurate
structures formed by arbitrary materials and with arbitrary lattice
mismatch/misalignment.

The structure of this paper is as follows. In Section~\ref{sec:Hamiltonian},
we review the description of vdW structures based on tight-binding
models and generalized umklapp processes developed in Refs.~\citep{Bistritzer_2010,Koshino_2015}.
We use this description of vdW structures to compute the ARPES matrix
elements for an arbitrary structure in Section~\ref{sec:ARPES-in-vdW}.
We apply the general framework to model ARPES in twisted bilayer graphene
and twisted bilayer MoS$_{2}$ in Section~\ref{sec:Applications}.
Conclusions are drawn in Section~\ref{sec:Conclusions}. For completeness
and the convenience of the reader, in Appendix~\ref{appx:ARPES_derivation},
we briefly review the derivation of the ARPES intensity within the
non-equilibrium Green's function approach \citep{Caroli_1973,Feibelman_1974}.
In Appendix~\ref{appx:interlayer_coupling}, we present some details
on the evaluation of the Fourier transform components of the interlayer
hopping in twisted bilayer MoS$_{2}$.

\section{Tight-binding description of vdW structures\label{sec:Hamiltonian}}

A first step in modeling ARPES of vdW structures is to describe the
electronic states bound to the structure. Following Refs.~\citep{Bistritzer_2010,Koshino_2015},
we employ a tight-binding model to describe the bound states. We will
focus on bilayer structures, where each layer has a periodic structure
with Bravais lattice sites given by $\left\{ \mathbf{R}_{\ell,i}\right\} $,
with $\ell=t,b$ labeling the top and bottom layers, respectively.
The single-particle Hamiltonian of the structure is written as
\begin{equation}
H=H_{t}+H_{b}+H_{tb}+H_{bt},\label{eq:Hamiltonian_bilayer}
\end{equation}
where $H_{t}$ and $H_{b}$ are the tight-binding Hamiltonians of
the isolated top and bottom layers and $H_{tb}$ ($H_{bt}$) describes
the hopping of electrons from the bottom (top) to the top (bottom)
layer. More concretely, the intralayer terms are written as 
\begin{equation}
H_{\ell}=\sum_{\substack{i,j,\alpha,\beta}
}h_{\alpha\beta}^{\ell,\ell}\left(\mathbf{R}_{\ell,i},\mathbf{R}_{\ell,j}\right)c_{\ell,\mathbf{R}_{\ell,i},\alpha}^{\dagger}c_{t,\mathbf{R}_{\ell,j},\beta}
\end{equation}
for $\ell=t,b$, and the interlayer terms as 
\begin{equation}
H_{tb}=\sum_{\substack{i,j,\alpha,\beta}
}h_{\alpha\beta}^{t,b}\left(\mathbf{R}_{t,i},\mathbf{R}_{b,j}\right)c_{t,\mathbf{R}_{t,i},\alpha}^{\dagger}c_{b,\mathbf{R}_{b,j},\beta},
\end{equation}
with $H_{bt}=H_{tb}^{\dagger}$. In the previous equations, the indices
$i,j$ run over lattice sites and the indices $\alpha,\beta$ run
over sublattice, orbital and spin degrees of freedom. The operator
$c_{\ell,\mathbf{R}_{\ell,i},\alpha}^{\dagger}$ creates an electron
in state $\left|\ell,\mathbf{R}_{\ell,i},\alpha\right\rangle $, a
localized Wannier state in layer $\ell$, lattice site $\mathbf{R}_{\ell,i}$
and sublattice site $\bm{\tau}_{\ell,\alpha}$. The Wannier wavefunction
in real space reads $\left\langle \mathbf{r}\left|\ell,\mathbf{R}_{\ell,i},\alpha\right.\right\rangle =w_{\ell,\alpha}\left(\mathbf{r}-\mathbf{R}_{\ell,i}-\bm{\tau}_{\ell,\alpha}\right)$
where $w_{\ell,\alpha}\left(\mathbf{r}\right)$ is the Wannier wavefunction
or type $\alpha$ centered on the origin. $h_{\alpha\beta}^{t,t}\left(\mathbf{R}_{t,i},\mathbf{R}_{t,j}\right)$
and $h_{\alpha\beta}^{b,b}\left(\mathbf{R}_{b,i},\mathbf{R}_{b,j}\right)$
are intralayer hopping terms, which we assume to be invariant under
translations by lattice vector of the respective layer. $h_{\alpha\beta}^{t,b}\left(\mathbf{R}_{t,i},\mathbf{R}_{b,j}\right)$
and $h_{\alpha\beta}^{b,t}\left(\mathbf{R}_{b,i},\mathbf{R}_{t,j}\right)$
are interlayer hopping terms, which describe the coupling between
the two layers. It is convenient to express the electronic operators
in terms of Fourier components 
\begin{equation}
c_{\ell,\mathbf{R}_{\ell,i},\alpha}^{\dagger}=\frac{1}{\sqrt{N_{\ell}}}\sum_{\mathbf{k}_{\ell}}e^{-i\mathbf{k}_{\ell}\cdot\left(\mathbf{R}_{\ell,i}+\bm{\tau}_{\ell,\alpha}\right)}c_{\ell,\mathbf{k}_{\ell},\alpha}^{\dagger},\label{eq:Bloch-Wannier_states}
\end{equation}
where $\mathbf{k}_{\ell}$ belongs to the Brillouin zone of layer
$\ell$ and $N_{\ell}$ is the number of unit cells in layer $\ell$.
Notice that if $\mathbf{G}_{\ell}$ is a reciprocal lattice vector
of layer $\ell$, i.e., $e^{i\mathbf{G}_{\ell}\cdot\mathbf{R}_{\ell,i}}=1$,
then $c_{\ell,\mathbf{k}+\mathbf{G}_{\ell},\alpha}^{\dagger}=e^{i\mathbf{G}_{\ell}\cdot\bm{\tau}_{\ell,\alpha}}c_{\ell,\mathbf{k},\alpha}^{\dagger}$.
These states bring the Hamiltonians of the isolated layers to a block
diagonal form, 
\begin{equation}
H_{\ell}=\sum_{\mathbf{k}_{\ell},\alpha,\beta}c_{\ell,\mathbf{k}_{\ell},\alpha}^{\dagger}h_{\alpha\beta}^{\ell,\ell}\left(\mathbf{k}_{\ell}\right)c_{\ell,\mathbf{k}_{\ell},\beta},
\end{equation}
where $h_{\alpha\beta}^{\ell,\ell}\left(\mathbf{k}_{\ell}\right)=\sum_{i}e^{-i\mathbf{k}_{\ell}\cdot\left(\mathbf{R}_{\ell,i}+\bm{\tau}_{\ell,\alpha}-\bm{\tau}_{\ell,\beta}\right)}h_{\alpha\beta}^{\ell,\ell}\left(\mathbf{R}_{\ell,i},\mathbf{0}\right)$.
For the interlayer term, we assume a two-center approximation for
the hopping elements and write them in terms of their Fourier transform
components \citep{Koshino_2015} as (focusing on the $H_{tb}$ term)
\begin{multline}
h_{\alpha\beta}^{t,b}\left(\mathbf{R}_{t,i},\mathbf{R}_{b,j}\right)=\sqrt{A_{\text{c},t}A_{\text{c},b}}\times\\
\times\int\frac{d^{2}\mathbf{q}}{\left(2\pi\right)^{2}}e^{i\mathbf{q}\cdot\left(\mathbf{R}_{t,i}+\bm{\tau}_{t,\alpha}-\mathbf{R}_{b,j}-\bm{\tau}_{b,\beta}\right)}h_{\alpha\beta}^{t,b}\left(\mathbf{q}\right),
\end{multline}
where $A_{\text{c},\ell}$ is the area of the unit cell of layer $\ell$.
With this we can write $H_{tb}$ as 
\begin{multline}
H_{tb}=\sum_{\mathbf{k}_{t},\mathbf{k}_{b},\alpha,\beta}\sum_{i,j}e^{i\bm{\tau}_{t,\alpha}\cdot\mathbf{G}_{t,i}}h_{\alpha\beta}^{t,b}\left(\mathbf{k}_{t}+\mathbf{G}_{t,i}\right)e^{-i\bm{\tau}_{b,\beta}\cdot\mathbf{G}_{b,j}}\times\\
\times c_{t,\mathbf{k}_{t}+\mathbf{G}_{t,i},\alpha}^{\dagger}c_{b,\mathbf{k}_{b}+\mathbf{G}_{b,j},\beta}\delta_{\mathbf{k}_{t}+\mathbf{G}_{t,i},\mathbf{k}_{b}+\mathbf{G}_{b,j}},\label{eq:umklapp_Hamiltonian}
\end{multline}
where $\mathbf{G}_{\ell,i}$ are reciprocal lattice vectors of layer
$\ell$. $H_{bt}$ is written in a similar way.

Equation~(\ref{eq:umklapp_Hamiltonian}) tells us that states of
the two layers with crystal momentum $\mathbf{k}_{t}$ and $\mathbf{k}_{b}$
are only coupled provided $\mathbf{G}_{t,i}$ and $\mathbf{G}_{b,j}$
exist, such that the generalized umklapp condition $\mathbf{k}_{t}+\mathbf{G}_{t,i}=\mathbf{k}_{b}+\mathbf{G}_{b,j}$
is satisfied \citep{Koshino_2015}. In an extended Brillouin zone
scheme, the generalized umklapp condition can be satisfied if for
each $\mathbf{G}_{t,i}$ and $\mathbf{G}_{b,j}$ we write $\mathbf{k}_{t}=\mathbf{k}+\mathbf{G}_{b,j}$
and $\mathbf{k}_{b}=\mathbf{k}+\mathbf{G}_{t,i}$ for any $\mathbf{k}$.
This fact motivates us to look for eigenstates of the complete Hamiltonian
Eq.~(\ref{eq:Hamiltonian_bilayer}) of the form 
\begin{align}
\left|\psi_{\mathbf{k},n}^{\text{vdW}}\right\rangle  & =\sum_{i,\alpha}\phi_{t,\mathbf{k},\alpha}^{n}\left(\mathbf{G}_{b,i}\right)c_{t,\mathbf{k}+\mathbf{G}_{b,i},\alpha}^{\dagger}\left|0\right\rangle \nonumber \\
 & +\sum_{i,\alpha}\phi_{b,\mathbf{k},\alpha}^{n}\left(\mathbf{G}_{t,i}\right)c_{b,\mathbf{k}+\mathbf{G}_{t,i},\alpha}^{\dagger}\left|0\right\rangle ,\label{eq:eigenstate_expansion}
\end{align}
with $\phi_{t,\mathbf{k},\alpha}^{n}\left(\mathbf{G}_{b,i}\right)$
and $\phi_{b,\mathbf{k},\alpha}^{n}\left(\mathbf{G}_{t,i}\right)$
coefficients that are to be determined. We now introduce a convenient
compact notation. We define $\bm{\phi}_{\ell,\mathbf{k}}^{n}\left(\mathbf{G}\right)$
as a vector with entries given by $\left[\bm{\phi}_{\ell,\mathbf{k}}^{n}\left(\mathbf{G}\right)\right]_{\alpha}=$
$\phi_{\ell,\mathbf{k},\alpha}^{n}\left(\mathbf{G}\right)$. In the
same way we introduce the matrices $\bm{h}_{\mathbf{k}}^{\ell,\ell}$
with entries $\left[\bm{h}_{\mathbf{k}}^{\ell,\ell}\right]_{\alpha\beta}=h_{\alpha\beta}^{\ell,\ell}\left(\mathbf{k}\right)$,
$\bm{h}_{\mathbf{k},\mathbf{G}_{b},\mathbf{G}_{t}}^{t,b}$ with entries
$\left[\bm{h}_{\mathbf{k},\mathbf{G}_{b},\mathbf{G}_{t}}^{t,b}\right]_{\alpha\beta}=e^{i\mathbf{G}_{t}\cdot\bm{\tau}_{t,\alpha}}h_{\alpha\beta}^{t,b}\left(\mathbf{k}+\mathbf{G}_{b}+\mathbf{G}_{t}\right)e^{-i\mathbf{G}_{b}\cdot\bm{\tau}_{b,\beta}}$
and $\bm{h}_{\mathbf{k},\mathbf{G}_{t},\mathbf{G}_{b}}^{b,t}$ defined
in a similar way. This allows us to write the eigenvalue problem which
determines the eigenstates and energies of the vdW structure as 
\begin{equation}
\bm{H}_{\mathbf{k}}\left(\left\{ \mathbf{G}_{b}\right\} ,\left\{ \mathbf{G}_{t}\right\} \right)\cdot\left[\begin{array}{c}
\bm{\phi}_{t,\mathbf{k}}^{n}\left(\left\{ \mathbf{G}_{b}\right\} \right)\\
\bm{\phi}_{b,\mathbf{k}}^{n}\left(\left\{ \mathbf{G}_{t}\right\} \right)
\end{array}\right]=E_{\mathbf{k},n}\left[\begin{array}{c}
\bm{\phi}_{t,\mathbf{k}}^{n}\left(\left\{ \mathbf{G}_{b}\right\} \right)\\
\bm{\phi}_{b,\mathbf{k}}^{n}\left(\left\{ \mathbf{G}_{t}\right\} \right)
\end{array}\right],\label{eq:eigenvector_incommensurate}
\end{equation}
where $E_{\mathbf{k},n}$ are the energies, 
\begin{align}
\bm{\phi}_{t,\mathbf{k}}^{n}\left(\left\{ \mathbf{G}_{b}\right\} \right) & =\left[\begin{array}{ccc}
\bm{\phi}_{t,\mathbf{k}}^{n}\left(\mathbf{G}_{b,1}\right) & \bm{\phi}_{t,\mathbf{k}}^{n}\left(\mathbf{G}_{b,2}\right) & \cdots\end{array}\right]^{t},\\
\bm{\phi}_{b,\mathbf{k}}^{n}\left(\left\{ \mathbf{G}_{t}\right\} \right) & =\left[\begin{array}{ccc}
\bm{\phi}_{b,\mathbf{k}}^{n}\left(\mathbf{G}_{t,1}\right) & \bm{\phi}_{b,\mathbf{k}}^{n}\left(\mathbf{G}_{t,2}\right) & \cdots\end{array}\right]^{t},
\end{align}
are vectors formed by the coefficients $\phi_{t,\mathbf{k},\alpha}^{n}\left(\mathbf{G}_{b,i}\right)$
and $\phi_{b,\mathbf{k},\alpha}^{n}\left(\mathbf{G}_{t,i}\right)$
for different $\mathbf{G}_{b,i}$ and $\mathbf{G}_{t,i}$, and the
Hamiltonian matrix is written as 
\begin{equation}
\bm{H}_{\mathbf{k}}\left(\left\{ \mathbf{G}_{b}\right\} ,\left\{ \mathbf{G}_{t}\right\} \right)=\left[\begin{array}{cc}
\bm{\mathcal{H}}_{\mathbf{k}+\left\{ \mathbf{G}_{b}\right\} }^{t,t} & \bm{\mathcal{H}}_{\mathbf{k},\left\{ \mathbf{G}_{b}\right\} ,\left\{ \mathbf{G}_{t}\right\} }^{t,b}\\
\bm{\mathcal{H}}_{\mathbf{k},\left\{ \mathbf{G}_{t}\right\} ,\left\{ \mathbf{G}_{b}\right\} }^{b,t} & \bm{\mathcal{H}}_{\mathbf{k}+\left\{ \mathbf{G}_{t}\right\} }^{b,b}
\end{array}\right],\label{eq:Hamiltonian_vdW_full}
\end{equation}
where $\bm{\mathcal{H}}_{\mathbf{k}+\left\{ \mathbf{G}_{b}\right\} }^{t,t}$
is a block diagonal matrix 
\begin{equation}
\bm{\mathcal{H}}_{\mathbf{k}+\left\{ \mathbf{G}_{b}\right\} }^{t,t}=\left[\begin{array}{ccc}
\bm{h}_{\mathbf{k}+\mathbf{G}_{b,1}}^{t,t} & \bm{0} & \cdots\\
\bm{0} & \bm{h}_{\mathbf{k}+\mathbf{G}_{b,2}}^{t,t}\\
\vdots &  & \ddots
\end{array}\right],\label{eq:Hamiltonian_vdW_top}
\end{equation}
with $\bm{\mathcal{H}}_{\mathbf{k}+\left\{ \mathbf{G}_{t}\right\} }^{b,b}$
similarly defined, and $\bm{\mathcal{H}}_{\mathbf{k},\left\{ \mathbf{G}_{b}\right\} ,\left\{ \mathbf{G}_{t}\right\} }^{t,b}$
is a dense matrix

\begin{equation}
\bm{\mathcal{H}}_{\mathbf{k},\left\{ \mathbf{G}_{b}\right\} ,\left\{ \mathbf{G}_{t}\right\} }^{t,b}=\left[\begin{array}{ccc}
\bm{h}_{\mathbf{k},\mathbf{G}_{b,1},\mathbf{G}_{t,1}}^{t,b} & \bm{h}_{\mathbf{k},\mathbf{G}_{b,1},\mathbf{G}_{t,2}}^{t,b} & \cdots\\
\bm{h}_{\mathbf{k},\mathbf{G}_{b,2},\mathbf{G}_{t,1}}^{t,b} & \bm{h}_{\mathbf{k},\mathbf{G}_{b,2},\mathbf{G}_{t,2}}^{t,b} & \cdots\\
\vdots & \vdots & \ddots
\end{array}\right],\label{eq:Hamiltonian_vdW_top_bottom}
\end{equation}
with $\bm{\mathcal{H}}_{\mathbf{k},\left\{ \mathbf{G}_{t}\right\} ,\left\{ \mathbf{G}_{b}\right\} }^{b,t}=\left(\bm{\mathcal{H}}_{\mathbf{k},\left\{ \mathbf{G}_{b}\right\} ,\left\{ \mathbf{G}_{t}\right\} }^{t,b}\right)^{\dagger}$.
For a commensurate structure, there exist $\mathbf{G}_{b,i}$ and
$\mathbf{G}_{t,j}$ such that $\mathbf{G}_{b,i}=\mathbf{G}_{t,j}$
and the sums over reciprocal lattice vectors in Eq.~(\ref{eq:eigenstate_expansion})
become finite, and consequently the matrix $\bm{H}_{\mathbf{k}}\left(\left\{ \mathbf{G}_{b}\right\} ,\left\{ \mathbf{G}_{t}\right\} \right)$
is finite. In this case, the eigenvalues and eigenstates of $\bm{H}_{\mathbf{k}}\left(\left\{ \mathbf{G}_{b}\right\} ,\left\{ \mathbf{G}_{t}\right\} \right)$
for $\mathbf{k}$ restricted to the Brillouin zone of the commensurate
structure provide us the full spectrum and eigenstates of the bilayer
structure. For an incommensurate structure, the sums over reciprocal
lattice vectors in Eq.~(\ref{eq:eigenstate_expansion}) involve an
infinite number of terms and the corresponding Hamiltonian matrix
$\bm{H}_{\mathbf{k}}\left(\left\{ \mathbf{G}_{b}\right\} ,\left\{ \mathbf{G}_{t}\right\} \right)$
is infinite. However, an approximation to the eigenstates and energies
of the incommensurate structure can still be obtained by suitably
truncating $\bm{H}_{\mathbf{k}}\left(\left\{ \mathbf{G}_{b}\right\} ,\left\{ \mathbf{G}_{t}\right\} \right)$,
considering a finite number of reciprocal lattice vectors $\mathbf{G}_{t,i}$
and $\mathbf{G}_{b,i}$. Even in the case of a commensurate structure,
the supercell might be very large, leading to a very large matrix
$\bm{H}_{\mathbf{k}}\left(\left\{ \mathbf{G}_{b}\right\} ,\left\{ \mathbf{G}_{t}\right\} \right)$,
and in that situation it might still be beneficial to compute the
eigenstates and energies of the system approximately by truncating
$\bm{H}_{\mathbf{k}}\left(\left\{ \mathbf{G}_{b}\right\} ,\left\{ \mathbf{G}_{t}\right\} \right)$.
In the following, we will generally refer to the approximate eigenvalues
$E_{\mathbf{k},n}$ as band structure, even in the case where we have
an incommensurate structure and the concept of Brillouin zone no longer
applies.

We will now prove a formal relation satisfied by the solutions of
Eq.~(\ref{eq:eigenvector_incommensurate}) that will be useful in
the next section. Assume that 
\begin{equation}
\left[\begin{array}{c}
\bm{\phi}_{t,\mathbf{k}}^{n}\left(\left\{ \mathbf{G}_{b}\right\} \right)\\
\bm{\phi}_{b,\mathbf{k}}^{n}\left(\left\{ \mathbf{G}_{t}\right\} \right)
\end{array}\right],
\end{equation}
is an eigenstate of $\bm{H}_{\mathbf{k}}\left(\left\{ \mathbf{G}_{b}\right\} ,\left\{ \mathbf{G}_{t}\right\} \right)$
with eigenvalue $E_{\mathbf{k},n}$. Then 
\begin{equation}
\left[\begin{array}{c}
e^{-i\mathbf{G}_{t,j}\cdot\bm{\tau}_{t}}\bm{\phi}_{t,\mathbf{k}}^{n}\left(\left\{ \mathbf{G}_{b}\right\} \right)\\
\bm{\phi}_{b,\mathbf{k}}^{n}\left(\left\{ \mathbf{G}_{t,j}+\mathbf{G}_{t}\right\} \right)
\end{array}\right],
\end{equation}
is an eigenstate of $\bm{H}_{\mathbf{k}+\mathbf{G}_{t,j}}\left(\left\{ \mathbf{G}_{b}\right\} ,\left\{ \mathbf{G}_{t}\right\} \right)$
with the same eigenvalue. This allows us to identify 
\begin{align}
\phi_{t,\mathbf{k}+\mathbf{G}_{t,j},\alpha}^{n}\left(\mathbf{G}_{b}\right) & =e^{-i\mathbf{G}_{t,j}\cdot\bm{\tau}_{t,\alpha}}\phi_{t,\mathbf{k},\alpha}^{n}\left(\mathbf{G}_{b}\right),\label{eq:top_translation_by_G_top}\\
\phi_{b,\mathbf{k}+\mathbf{G}_{t,j},\alpha}^{n}\left(\mathbf{G}_{t}\right) & =\phi_{b,\mathbf{k},\alpha}^{n}\left(\mathbf{G}_{t,j}+\mathbf{G}_{t}\right).\label{eq:bottom_translation_by_G_top}
\end{align}
This statement can be proved by looking at the structure of $\bm{H}_{\mathbf{k}+\mathbf{G}_{t,j}}\left(\left\{ \mathbf{G}_{b}\right\} ,\left\{ \mathbf{G}_{t}\right\} \right)$.
First, we notice that 
\begin{align}
\left[\bm{h}_{\mathbf{k}+\mathbf{G}_{t,j}+\mathbf{G}_{b,i}}^{t,t}\right]_{\alpha\beta} & =e^{-i\mathbf{G}_{t,j}\cdot\bm{\tau}_{t,\alpha}}\left[\bm{h}_{\mathbf{k}+\mathbf{G}_{b,i}}^{t,t}\right]_{\alpha\beta}e^{i\mathbf{G}_{t,j}\cdot\bm{\tau}_{t,\beta}},\\
\left[\bm{h}_{\mathbf{k}+\mathbf{G}_{t,j},\mathbf{G}_{b},\mathbf{G}_{t}}^{t,b}\right]_{\alpha\beta} & =e^{-i\mathbf{G}_{t,j}\cdot\bm{\tau}_{t,\alpha}}\left[\bm{h}_{\mathbf{k},\mathbf{G}_{b},\mathbf{G}_{t}+\mathbf{G}_{t,j}}^{t,b}\right]_{\alpha\beta}.
\end{align}
With these relations, we can write 
\begin{multline}
\bm{H}_{\mathbf{k}+\mathbf{G}_{t,j}}\left(\left\{ \mathbf{G}_{b}\right\} ,\left\{ \mathbf{G}_{t}\right\} \right)=\left[\begin{array}{cc}
e^{-i\mathbf{G}_{t,j}\cdot\bm{\tau}_{t}} & \bm{0}\\
\bm{0} & \bm{1}
\end{array}\right]\\
\cdot\bm{H}_{\mathbf{k}}\left(\left\{ \mathbf{G}_{b}\right\} ,\left\{ \mathbf{G}_{t}+\mathbf{G}_{t,j}\right\} \right)\cdot\left[\begin{array}{cc}
e^{i\mathbf{G}_{t,j}\cdot\bm{\tau}_{t}} & \bm{0}\\
\bm{0} & \bm{1}
\end{array}\right].
\end{multline}
For a commensurate structure, the set of vectors $\left\{ \mathbf{G}_{t}\right\} $
is finite and periodic modulo reciprocal lattice vectors of the commensurate
lattice. Therefore $\left\{ \mathbf{G}_{t}+\mathbf{G}_{t,j}\right\} $
coincides with $\left\{ \mathbf{G}_{t}\right\} $ apart from a reordering
of the vectors. Assume that this reordering is implemented by a permutation
matrix $\bm{P}$ such that $\bm{P}\left\{ \mathbf{G}_{t}\right\} =\left\{ \mathbf{G}_{t}+\mathbf{G}_{t,j}\right\} $.
Since both the Hamiltonian $\bm{H}_{\mathbf{k}}\left(\left\{ \mathbf{G}_{b}\right\} ,\left\{ \mathbf{G}_{t}+\mathbf{G}_{t,j}\right\} \right)$
and the vector $\left[\begin{array}{cc}
\bm{\phi}_{t,\mathbf{k}}^{n}\left(\left\{ \mathbf{G}_{b}\right\} \right) & \bm{\phi}_{b,\mathbf{k}}^{n}\left(\left\{ \mathbf{G}_{t,j}+\mathbf{G}_{t}\right\} \right)\end{array}\right]^{t}$ are reordered in the same way, we can write
\begin{widetext}
\begin{multline}
\bm{H}_{\mathbf{k}+\mathbf{G}_{t,j}}\left(\left\{ \mathbf{G}_{b}\right\} ,\left\{ \mathbf{G}_{t}\right\} \right)\cdot\left[\begin{array}{c}
e^{-i\mathbf{G}_{t,j}\cdot\bm{\tau}_{t}}\bm{\phi}_{t,\mathbf{k}}^{n}\left(\left\{ \mathbf{G}_{b}\right\} \right)\\
\bm{\phi}_{b,\mathbf{k}}^{n}\left(\left\{ \mathbf{G}_{t,j}+\mathbf{G}_{t}\right\} \right)
\end{array}\right]=\left[\begin{array}{cc}
e^{-i\mathbf{G}_{t,j}\cdot\bm{\tau}_{t}} & \bm{0}\\
\bm{0} & \bm{1}
\end{array}\right]\cdot\bm{H}_{\mathbf{k}}\left(\left\{ \mathbf{G}_{b}\right\} ,\left\{ \mathbf{G}_{t}+\mathbf{G}_{t,j}\right\} \right)\cdot\left[\begin{array}{c}
\bm{\phi}_{t,\mathbf{k}}^{n}\left(\left\{ \mathbf{G}_{b}\right\} \right)\\
\bm{\phi}_{b,\mathbf{k}}^{n}\left(\left\{ \mathbf{G}_{t,j}+\mathbf{G}_{t}\right\} \right)
\end{array}\right]\\
=\left[\begin{array}{cc}
e^{-i\mathbf{G}_{t,j}\cdot\bm{\tau}_{t}} & \bm{0}\\
\bm{0} & \bm{1}
\end{array}\right]\cdot\bm{P}\cdot\bm{P}^{-1}\cdot\bm{H}_{\mathbf{k}}\left(\left\{ \mathbf{G}_{b}\right\} ,\left\{ \mathbf{G}_{t}+\mathbf{G}_{t,j}\right\} \right)\cdot\bm{P}\cdot\bm{P}^{-1}\cdot\left[\begin{array}{c}
\bm{\phi}_{t,\mathbf{k}}^{n}\left(\left\{ \mathbf{G}_{b}\right\} \right)\\
\bm{\phi}_{b,\mathbf{k}}^{n}\left(\left\{ \mathbf{G}_{t,j}+\mathbf{G}_{t}\right\} \right)
\end{array}\right]\\
=\left[\begin{array}{cc}
e^{-i\mathbf{G}_{t,j}\cdot\bm{\tau}_{t}} & \bm{0}\\
\bm{0} & \bm{1}
\end{array}\right]\cdot\bm{P}\cdot\bm{H}_{\mathbf{k}}\left(\left\{ \mathbf{G}_{b}\right\} ,\left\{ \mathbf{G}_{t}\right\} \right)\cdot\left[\begin{array}{c}
\bm{\phi}_{t,\mathbf{k}}^{n}\left(\left\{ \mathbf{G}_{b}\right\} \right)\\
\bm{\phi}_{b,\mathbf{k}}^{n}\left(\left\{ \mathbf{G}_{t}\right\} \right)
\end{array}\right]\\
=E_{\mathbf{k},n}\left[\begin{array}{cc}
e^{-i\mathbf{G}_{t,j}\cdot\bm{\tau}_{t}} & \bm{0}\\
\bm{0} & \bm{1}
\end{array}\right]\cdot\bm{P}\cdot\left[\begin{array}{c}
\bm{\phi}_{t,\mathbf{k}}^{n}\left(\left\{ \mathbf{G}_{b}\right\} \right)\\
\bm{\phi}_{b,\mathbf{k}}^{n}\left(\left\{ \mathbf{G}_{t}\right\} \right)
\end{array}\right]=E_{\mathbf{k},n}\left[\begin{array}{c}
e^{-i\mathbf{G}_{t,j}\cdot\bm{\tau}_{t}}\bm{\phi}_{t,\mathbf{k}}^{n}\left(\left\{ \mathbf{G}_{b}\right\} \right)\\
\bm{\phi}_{b,\mathbf{k}}^{n}\left(\left\{ \mathbf{G}_{t,j}+\mathbf{G}_{t}\right\} \right)
\end{array}\right],\label{eq:reordered_Hamiltonian}
\end{multline}
\end{widetext}

proving our statement for the commensurate case. For an incommensurate
structure, the set $\left\{ \mathbf{G}_{t}\right\} $ is infinite
and therefore, apart from a reordering, we have that $\left\{ \mathbf{G}_{t,j}+\mathbf{G}_{t}\right\} $
and $\left\{ \mathbf{G}_{t}\right\} $ coincide\footnote{Notice that this is only true formally for the incommensurate case,
not being true for any finite truncation of the matrix $\bm{H}_{\mathbf{k}}\left(\left\{ \mathbf{G}_{b}\right\} ,\left\{ \mathbf{G}_{t}\right\} \right)$ } and we also obtain Eq.~(\ref{eq:reordered_Hamiltonian}). This proves
our formal statement for both the commensurate and the incommensurate
cases. Following the same argumentation it can also be formally shown
that 
\begin{align}
\phi_{t,\mathbf{k}+\mathbf{G}_{b,j},\alpha}^{n}\left(\mathbf{G}_{b}\right) & =\phi_{t,\mathbf{k},\alpha}^{n}\left(\mathbf{G}_{b}+\mathbf{G}_{b,j}\right),\label{eq:top_translation_by_G_bottom}\\
\phi_{b,\mathbf{k}+\mathbf{G}_{b,j},\alpha}^{n}\left(\mathbf{G}_{t}\right) & =e^{-i\mathbf{G}_{b,j}\cdot\bm{\tau}_{t,\alpha}}\phi_{b,\mathbf{k},\alpha}^{n}\left(\mathbf{G}_{t}\right).\label{eq:bottom_translation_by_G_bottom}
\end{align}

In this section, the coupling between the two layers was assumed to
only give origin to interlayer hopping terms, not affecting the intralayer
Hamiltonians $H_{t}$ and $H_{b}$, which were assumed to preserve
the translational symmetry of the isolated layers. Besides this effect,
there is also the possibility of one of the layers inducing a potential
to which the electrons in the other layer will be subjected to \citep{Ortix_2012,Yankowitz_2012,Wallbank_2013}.
The coupling between the layers can also lead to structural relaxation,
which gives origin to a modulation of the intralayer hoppings due
to the displacement of the atomic positions \citep{SanJose_2014,Slotman_2015,Jung_2015,Jung_2017}.
Although we will not explore those effects in the present work, we
note that these corrections will have a spatial modulation given by
$\mathbf{G}_{t,i}-\mathbf{G}_{b,j}$ and can therefore be incorporated
in the present formalism by including off-diagonal blocks in the matrices
$\bm{\mathcal{H}}_{\mathbf{k}+\left\{ \mathbf{G}_{b}\right\} }^{t,t}$
and $\bm{\mathcal{H}}_{\mathbf{k}+\left\{ \mathbf{G}_{b}\right\} }^{b,b}$
of the form 
\begin{multline}
\left[\bm{V}_{\mathbf{k}+\mathbf{G}_{b(t),i},\mathbf{k}+\mathbf{G}_{b(t),j}}^{t,t(b,b)}\right]_{\alpha\beta}=\\
=\left\langle t(b),\mathbf{k}+\mathbf{G}_{b(t),i},\alpha\right|V^{t(b)}\left|t(b),\mathbf{k}+\mathbf{G}_{b(t),j},\beta\right\rangle ,
\end{multline}
where $V^{t(b)}$ describes the potential or intralayer hopping modulation
on the top (bottom) layer and $\left|\ell,\mathbf{k},\alpha\right\rangle $
is the state created by the operator $c_{\ell,\mathbf{k},\alpha}^{\dagger}$
in Eq.~\eqref{eq:Bloch-Wannier_states}.

\section{ARPES in vdW structures\label{sec:ARPES-in-vdW}}

We wish to model an experimental situation where the incident electromagnetic
field is monochromatic with frequency $\omega_{0}>0$, and the electron
detector is placed at position $\mathbf{r}$, far away from the crystal
sample, collecting electrons emitted with energy $E$ along the direction
$\hat{\mathbf{r}}$. In this situation, the energy resolved ARPES
intensity can be evaluated from \citep{Adawi_1964,Mahan_1970,Schaich_1970,Schaich_1971,Caroli_1973,Feibelman_1974,Pendry_1976}
(see also Appendix~\ref{appx:ARPES_derivation} for a brief derivation)
\begin{multline}
I_{\text{ARPES}}(E,\hat{\mathbf{r}})\propto\sum_{a}f\left(E-\omega_{0}-\mu\right)\times\\
\times\left|M_{E,\hat{\mathbf{r}};a}(\omega_{0})\right|^{2}\frac{1}{2\pi}\mathcal{A}_{a}\left(E-\omega_{0}\right).\label{eq:ARPES_intensity}
\end{multline}
where $\mu$ is the chemical potential, the index $a$ runs over crystal
bound states, with corresponding wavefunction $\psi_{a}(\mathbf{r})$;
$\mathcal{A}_{a}\left(E\right)$ is the spectral function, which in
the non-interacting limit reduces to $\mathcal{A}_{a}\left(E\right)=2\pi\delta\left(E-\epsilon_{\alpha}\right)$,
where $\epsilon_{a}$ is the energy of state $a$; and $M_{E,\hat{\mathbf{r}};a}(\omega_{0})$
are the ARPES matrix elements. These are given by 
\begin{multline}
M_{E,\hat{\mathbf{r}};a}(\omega_{0})=-i\frac{e}{\hbar}\int d^{3}\mathbf{r}_{1}\left[\psi_{E,\hat{\mathbf{r}}}^{*}(\mathbf{r}_{1})\left(\nabla\psi_{a}(\mathbf{r}_{1})\right)\right.\\
\left.-\left(\nabla\psi_{E,\hat{\mathbf{r}}}^{*}(\mathbf{r}_{1})\right)\psi_{a}(\mathbf{r}_{1})\right]\cdot\mathbf{A}_{\omega_{0}}(\mathbf{r}_{1}),\label{eq:ARPES_matrix_elements}
\end{multline}
where $\mathbf{A}_{\omega_{0}}(\mathbf{r}_{1})$ is the screened \citep{Feibelman_1974}
vector potential and $\psi_{E,\hat{\mathbf{r}}}^{*}(\mathbf{r}_{1})$
is the complex conjugate of a photoemitted state with energy $E$,
which is the solution of the Lippmann-Schwinger equation 
\begin{equation}
\psi_{E,\hat{\mathbf{r}}}^{*}(\mathbf{r}_{1})=e^{-i\mathbf{p}_{E}\cdot\mathbf{r}_{1}}+\int d^{3}\mathbf{r}^{\prime}\psi_{E,\hat{\mathbf{r}}}^{*}(\mathbf{r}^{\prime})V(\mathbf{r}^{\prime})G_{\text{free}}^{R}(E;\mathbf{r}^{\prime},\mathbf{r}_{1})\label{eq:inverse_difraction_state}
\end{equation}
where $\mathbf{p}_{E}=p_{E}\hat{\mathbf{r}}$, with $p_{E}=\sqrt{2m\left(E+i0^{+}\right)/\hbar^{2}}$,
$G_{\text{free}}^{R}(E;\mathbf{r},\mathbf{r}^{\prime})$ is the free
space electronic Green's function (which is explicitly given by Eq.~\eqref{eq:free_space_GF}),
and $V(\mathbf{r})$ is the crystal potential. Using integration by
parts, we can write the ARPES matrix element as 
\begin{equation}
M_{E,\hat{\mathbf{r}};a}(\omega_{0})=i\frac{2e}{\hbar}\int d^{3}\mathbf{r}_{1}\left(\nabla\psi_{E,\hat{\mathbf{r}}}^{*}(\mathbf{r}_{1})\right)\psi_{a}(\mathbf{r}_{1})\cdot\mathbf{A}_{\omega_{0}}(\mathbf{r}_{1}),
\end{equation}
where we have neglected the contribution arising from $\nabla\cdot\mathbf{A}_{\omega_{0}}(\mathbf{r})=\rho/\left(i\omega\epsilon_{0}\right)$,
where $\rho$ is the total charge in the crystal, which is a common
approximation \citep{Feibelman_1974,Pendry_1976,Luders_2001}. In
the same spirit, we neglect effects of screening in $\mathbf{A}_{\omega_{0}}(\mathbf{r}_{1})$
and assume it to be described by a plane wave $\mathbf{A}_{\omega_{0}}(\mathbf{r}_{1})=A_{\omega_{0}}^{\lambda}\mathbf{e}_{\mathbf{q}}^{\lambda}e^{i\mathbf{q}\cdot\mathbf{r}_{1}}$,
where $A_{\omega_{0}}^{\lambda}$ is the amplitude, $\mathbf{q}$
is the wavevector, satisfying $\omega_{0}=c\left|\mathbf{q}\right|$
with $c$ the speed of light, and $\mathbf{e}_{\mathbf{q}}^{\lambda}$
is the polarization vector for the $\lambda=s,p$ polarizations. For
simplicity we will also approximate $\psi_{E,\hat{\mathbf{r}}}^{*}(\mathbf{r}_{1})$
by a plane wave $\psi_{E,\hat{\mathbf{r}}}^{*}(\mathbf{r}_{1})\simeq e^{-i\mathbf{p}_{E}\cdot\mathbf{r}_{1}}$\citep{Shirley_1995},
which will greatly simplify the evaluation of $M_{E,\hat{\mathbf{r}};a}(\omega_{0})$,
while providing a non-trivial description of the ARPES matrix elements.\footnote{This approximation is sometimes insufficient to explain the observed
data as shown in Ref.~\citep{Gierz_2011}. In this case, multiple
scatterings of the photoemitted electron by the lattice become important
and a full solution of the Lippmann-Schwinger equation (\ref{eq:inverse_difraction_state})
is required. } With these approximations we obtain 
\begin{equation}
M_{E,\hat{\mathbf{r}};a}(\omega_{0})\simeq\frac{2e}{\hbar}A_{\omega_{0}}^{\lambda}\left(\mathbf{p}_{E}\cdot\mathbf{e}_{\mathbf{q}}^{\lambda}\right)\int d^{3}\mathbf{r}_{1}e^{-i\left(\mathbf{p}_{E}-\mathbf{q}\right)\cdot\mathbf{r}_{1}}\psi_{a}(\mathbf{r}_{1})\label{eq:Matrix_element_plane_wave}
\end{equation}
and the ARPES matrix element becomes proportional to the Fourier transform
of the crystal bound state.

In order to obtain the ARPES intensity for a bilayer vdW structure
we need to evaluate Eq.~(\ref{eq:Matrix_element_plane_wave}) with
the crystal bound state given by Eq.~(\ref{eq:eigenstate_expansion}).
In real space we have that 

\begin{align}
\psi_{\mathbf{k},n}^{\text{vdW}}(\mathbf{r}) & =\frac{1}{\sqrt{N_{t}}}\sum_{i,j,\alpha}\phi_{t,\mathbf{k},\alpha}^{n}\left(\mathbf{G}_{b,i}\right)e^{i\left(\mathbf{k}+\mathbf{G}_{b,i}\right)\cdot\left(\mathbf{R}_{t,j}+\bm{\tau}_{t,\alpha}\right)}\nonumber \\
 & \times w_{t,\alpha}\left(\mathbf{r}-\mathbf{R}_{t,j}-\bm{\tau}_{t,\alpha}\right)\nonumber \\
 & +\frac{1}{\sqrt{N_{b}}}\sum_{i,j,\alpha}\phi_{b,\mathbf{k},\alpha}^{n}\left(\mathbf{G}_{t,i}\right)e^{i\left(\mathbf{k}+\mathbf{G}_{t,i}\right)\cdot\left(\mathbf{R}_{b,j}+\bm{\tau}_{b,\alpha}\right)}\nonumber \\
 & \times w_{b,\alpha}\left(\mathbf{r}-\mathbf{R}_{b,j}-\bm{\tau}_{b,\alpha}\right),\label{eq:wavefunction_real_space_vdW}
\end{align}
and the ARPES matrix element Eq.~(\ref{eq:Matrix_element_plane_wave})
becomes 
\begin{multline}
M_{E,\hat{\mathbf{r}};\mathbf{k},n}(\omega_{0})=\frac{2e}{\hbar}A_{\omega_{0}}^{\lambda}\left(\mathbf{p}_{E}\cdot\mathbf{e}_{\mathbf{q}}^{\lambda}\right)\sqrt{N_{t}}\\
\times\Biggl[\sum_{i,j,\alpha}\phi_{t,\mathbf{k},\alpha}^{n}\left(\mathbf{G}_{b,i}\right)e^{i\mathbf{G}_{t,j}\cdot\bm{\tau}_{t,\alpha}}e^{-iQ_{z}\tau_{t,\alpha}^{z}}\\
\times\tilde{w}_{t,\alpha}\left(\mathbf{Q}\right)\delta_{\mathbf{k}+\mathbf{G}_{b,i}-\mathbf{Q}_{\perp},\mathbf{G}_{t,j}}\\
+\sqrt{\frac{N_{b}}{N_{t}}}\sum_{i,j,\alpha}\phi_{b,\mathbf{k},\alpha}^{n}\left(\mathbf{G}_{t,i}\right)e^{i\mathbf{G}_{b,j}\cdot\bm{\tau}_{b,\alpha}}e^{-iQ_{z}\tau_{b,\alpha}^{z}}\\
\times\tilde{w}_{b,\alpha}\left(\mathbf{Q}\right)\delta_{\mathbf{k}+\mathbf{G}_{t,i}-\mathbf{Q}_{\perp},\mathbf{G}_{b,j}}\Biggr],
\end{multline}
where $\mathbf{Q}=\mathbf{p}_{E}-\mathbf{q}$ is the transferred momentum,
with $Q_{z}$ and $\mathbf{Q}_{\perp}$ indicating the components
parallel and perpendicular to the $z$ axis, and $\tilde{w}_{\ell,\alpha}\left(\mathbf{Q}\right)=\int d^{3}\mathbf{r}e^{-i\mathbf{Q}\cdot\mathbf{r}}w_{\ell,\alpha}(\mathbf{r})$
is the Fourier transform of the Wannier wave functions. Using the
relations given by Eqs.~(\ref{eq:top_translation_by_G_top}), (\ref{eq:bottom_translation_by_G_top}),
(\ref{eq:top_translation_by_G_bottom}) and (\ref{eq:bottom_translation_by_G_bottom})
together with the in-plane momentum-conserving Kronecker symbols,
we can write 
\begin{multline}
\phi_{t,\mathbf{k},\alpha}^{n}\left(\mathbf{G}_{b,i}\right)e^{i\mathbf{G}_{t,j}\cdot\bm{\tau}_{t,\alpha}}=\\
=\phi_{t,\mathbf{Q}_{\perp}+\mathbf{G}_{t,j}-\mathbf{G}_{b,i},\alpha}^{n}\left(\mathbf{G}_{b,i}\right)e^{i\mathbf{G}_{t,j}\cdot\bm{\tau}_{t,\alpha}}=\phi_{t,\mathbf{Q}_{\perp},\alpha}^{n}\left(\mathbf{0}\right),
\end{multline}
\begin{multline}
\phi_{b,\mathbf{k},\alpha}^{n}\left(\mathbf{G}_{t,i}\right)e^{i\mathbf{G}_{b,j}\cdot\bm{\tau}_{b,\alpha}}=\\
=\phi_{b,\mathbf{Q}_{\perp}+\mathbf{G}_{b,j}-\mathbf{G}_{t,i},\alpha}^{n}\left(\mathbf{G}_{t,i}\right)e^{i\mathbf{G}_{b,j}\cdot\bm{\tau}_{b,\alpha}}=\phi_{b,\mathbf{Q}_{\perp},\alpha}^{n}\left(\mathbf{0}\right),
\end{multline}
and the ARPES matrix elements can be rewritten as 
\begin{multline}
M_{E,\hat{\mathbf{r}};\mathbf{k},n}(\omega_{0})=\sqrt{N_{t}}\frac{2e}{\hbar}A_{\omega_{0}}^{\lambda}\left(\mathbf{p}_{E}\cdot\mathbf{e}_{\mathbf{q}}^{\lambda}\right)\mathcal{M}_{\mathbf{Q},n}\times\\
\times\sum_{i,j}\delta_{\mathbf{k}-\mathbf{Q}_{\perp},\mathbf{G}_{b,i}+\mathbf{G}_{t,j}},
\end{multline}
where we have defined 
\begin{align}
\mathcal{M}_{\mathbf{Q},n} & =\sum_{\alpha}\phi_{t,\mathbf{Q}_{\perp},\alpha}^{n}\left(\mathbf{0}\right)e^{-iQ_{z}\tau_{t,\alpha}^{z}}\tilde{w}_{t,\alpha}\left(\mathbf{Q}\right)\nonumber \\
 & +\sqrt{\frac{A_{\text{c},t}}{A_{\text{c},b}}}\sum_{\alpha}\phi_{b,\mathbf{Q}_{\perp},\alpha}^{n}\left(\mathbf{0}\right)e^{-iQ_{z}\tau_{b,\alpha}^{z}}\tilde{w}_{b,\alpha}\left(\mathbf{Q}\right),
\end{align}
and assumed that the total area of both layers is the same, that is
$N_{t}A_{\text{c},t}=N_{b}A_{\text{c},b}$ \citep{Koshino_2015}.
It is still necessary to evaluate $\tilde{w}_{\ell,\alpha}\left(\mathbf{Q}\right)$.
Assuming that the Wannier functions are well localized they can be
written in a separable form as \citep{Shirley_1995} 
\begin{equation}
w_{\ell,\alpha}\left(\mathbf{r}\right)=R_{\ell,\alpha}\left(\left|\mathbf{r}\right|\right)\mathcal{Y}_{l_{\alpha}}^{m_{\alpha}}\left(\hat{\mathbf{r}}\right),\label{eq:wannier_seperable}
\end{equation}
where $R_{\ell,\alpha}\left(\left|\mathbf{r}\right|\right)$ is the
radial wave function and $\mathcal{Y}_{l}^{m}\left(\hat{\mathbf{r}}\right)=N_{l}^{m}\left(-1\right)^{\left|m\right|}P_{l}^{\left|m\right|}\left(\cos\theta_{\hat{\mathbf{r}}}\right)\Phi_{m}\left(\phi_{\hat{\mathbf{r}}}\right)$
are real spherical harmonics, where 
\begin{equation}
N_{l}^{m}=\sqrt{\frac{\left(2l+1\right)}{2\pi}\frac{\left(l-\left|m\right|\right)!}{\left(l+\left|m\right|\right)!}},
\end{equation}
is a normalization factor, $P_{l}^{m}(x)$ is an associated Legendre
polynomial and $\Phi_{m}\left(\phi_{\hat{\mathbf{r}}}\right)$ is
defined as 
\begin{equation}
\Phi_{m}\left(\phi_{\hat{\mathbf{r}}}\right)=\begin{cases}
\cos\left(m\phi_{\hat{\mathbf{r}}}\right) & ,\,m>0\\
\frac{1}{\sqrt{2}} & ,\,m=0\\
\sin\left(\left|m\right|\phi_{\hat{\mathbf{r}}}\right) & ,\,m<0
\end{cases}.\label{eq:angular_inplane}
\end{equation}
Using the plane wave expansion \citep{NIST_book} 
\begin{equation}
e^{-i\mathbf{Q}\cdot\mathbf{r}}=4\pi\sum_{l=0}^{+\infty}\sum_{m=-l}^{l}\left(-i\right)^{l}j_{l}(Qr)\mathcal{Y}_{l}^{m}\left(\hat{\mathbf{Q}}\right)\mathcal{Y}_{l}^{m}\left(\hat{\mathbf{r}}\right),\label{eq:plane_wave_expansion}
\end{equation}
where $j_{l}(x)$ is a spherical Bessel function, we can write $\tilde{w}_{\ell,\alpha}\left(\mathbf{Q}\right)$
as 
\begin{equation}
\tilde{w}_{\ell,\alpha}\left(\mathbf{Q}\right)=\left(-i\right)^{l_{\alpha}}\mathcal{Y}_{l_{\alpha}}^{m_{\alpha}}\left(\hat{\mathbf{Q}}\right)\tilde{R}_{\ell,\alpha}\left(Q\right),
\end{equation}
where $\tilde{R}_{\ell,\alpha}\left(Q\right)=4\pi\int_{0}^{+\infty}drr^{2}j_{l_{\alpha}}\left(Qr\right)R_{\ell,\alpha}\left(r\right)$
and we have used the orthogonality property of the real spherical
harmonics. For the case in which $R_{\ell,\alpha}\left(r\right)$
are given by hydrogen-like wavefunctions 
\begin{equation}
R_{\ell,\alpha}\left(r\right)=\frac{2}{n_{\alpha}a_{*}^{3/2}}\mathcal{N}_{n_{\alpha},l_{\alpha}}e^{-\frac{1}{2}u}x^{l_{\alpha}}L_{n_{\alpha}-l_{\alpha}-1}^{l_{\alpha}+1}(x),
\end{equation}
where $\mathcal{N}_{n,l}=\sqrt{\left(n-l-1\right)!/\left(n+l\right)!}$,
$x=2r/(n_{\alpha}a_{*})$, $a_{*}=a_{0}/Z_{*}$ (with $a_{0}$ the
Bohr radius and $Z_{*}$ the effective nuclear charge), and $L_{n}^{\alpha}$
is a generalized Laguerre polynomial; then $\tilde{R}_{\ell,\alpha}\left(Q\right)$
can be evaluated analytically and is given by \citep{atoms_molecules_book,Moser_2017}
\begin{multline}
\tilde{R}_{\ell,\alpha}\left(Q\right)=4\pi a_{*}^{3/2}\mathcal{N}_{n_{\alpha},l_{\alpha}}n_{\alpha}^{2}2^{2l_{\alpha}+2}l_{\alpha}!\times\\
\times\frac{y^{l_{\alpha}}}{\left(y^{2}+1\right)^{l_{\alpha}+2}}C_{n_{\alpha}-l_{\alpha}-1}^{l_{\alpha}+1}\left(\frac{y^{2}-1}{y^{2}+1}\right),
\end{multline}
where $y=n_{\alpha}Q/a_{*}$ and $C_{n}^{\alpha}$ is a Gegenbauer
polynomial. Analytic expressions for $\tilde{R}_{\ell,\alpha}\left(Q\right)$
are also available if $R_{\ell,\alpha}\left(r\right)$ are approximated
by Slater type \citep{Dzevad_1989} or Gaussian type \citep{Niyazi_2016}
orbitals.

Summarizing the results of this section, we can write the ARPES intensity
corresponding to photoemitted electrons with energy $E>0$ emitted
along direction $\hat{\mathbf{r}}$, due to an incident electromagnetic
field with frequency $\omega_{0}$, wave number $\mathbf{q}$, and
polarization vector $\mathbf{e}_{\mathbf{q}}^{\lambda}$ as 
\begin{multline}
I_{\text{ARPES}}\left(E,\hat{\mathbf{r}}\mid\omega_{0},\mathbf{q},\lambda\right)\propto N_{t}\left|\frac{2e}{\hbar}A_{\omega_{0}}^{\lambda}\right|^{2}\left|\mathbf{p}_{E}\cdot\mathbf{e}_{\mathbf{q}}^{\lambda}\right|^{2}\times\\
\times\sum_{n}f\left(E-\omega_{0}-\mu\right)\left|\mathcal{M}_{\mathbf{Q},n}\right|^{2}\frac{1}{2\pi}\mathcal{A}_{\mathbf{Q}_{\perp},n}\left(E-\omega_{0}\right).\label{eq:ARPES_intensity_final}
\end{multline}
where, $\mathbf{p}_{E}=p_{E}\hat{\mathbf{r}}$ with $p_{E}=\sqrt{2mE/\hbar^{2}}$,
$\mathbf{Q}=\mathbf{p}_{E}-\mathbf{q}$, 
\begin{multline}
\mathcal{M}_{\mathbf{Q},n}=\\
=\sum_{\alpha}\phi_{t,\mathbf{Q}_{\perp},\alpha}^{n}\left(\mathbf{0}\right)e^{-iQ_{z}\tau_{t,\alpha}^{z}}\left(-i\right)^{l_{\alpha}}\mathcal{Y}_{l_{\alpha}}^{m_{\alpha}}\left(\hat{\mathbf{Q}}\right)\tilde{R}_{t,\alpha}\left(Q\right)+\\
+\sqrt{\frac{A_{\text{c},t}}{A_{\text{c},b}}}\sum_{\alpha}\phi_{b,\mathbf{Q}_{\perp},\alpha}^{n}\left(\mathbf{0}\right)e^{-iQ_{z}\tau_{b,\alpha}^{z}}\times\\
\times\left(-i\right)^{l_{\alpha}}\mathcal{Y}_{l_{\alpha}}^{m_{\alpha}}\left(\hat{\mathbf{Q}}\right)\tilde{R}_{b,\alpha}\left(Q\right),\label{eq:M_Q}
\end{multline}
and in order to include broadening effects we can approximate the
spectral function by a Lorentzian 
\begin{equation}
\mathcal{A}_{\mathbf{Q}_{\perp},n}\left(\omega\right)\simeq2\frac{\eta}{\left(\omega-E_{\mathbf{Q}_{\perp},n}\right)^{2}+\eta^{2}},
\end{equation}
with $\eta$ the broadening factor.

We make two remarks regarding Eq.~(\ref{eq:M_Q}). First, we point
out that its form depends on the chosen convention to define the Fourier
components of the electronic operators in Eq.~(\ref{eq:Bloch-Wannier_states}).
It is also possible to work with an alternative convention, where
\begin{equation}
c_{\ell,\mathbf{R}_{\ell,i},\alpha}^{\dagger}=\frac{1}{\sqrt{N_{\ell}}}\sum_{\mathbf{k}_{\ell}}e^{-i\mathbf{k}_{\ell}\cdot\mathbf{R}_{\ell,i}}\tilde{c}_{\ell,\mathbf{k}_{\ell},\alpha}^{\dagger}.\label{eq:alternative_Bloch_convention}
\end{equation}
The operators $c_{\ell,\mathbf{k}_{\ell},\alpha}^{\dagger}$ and $\tilde{c}_{\ell,\mathbf{k}_{\ell},\alpha}^{\dagger}$
are related via $\tilde{c}_{\ell,\mathbf{k}_{\ell},\alpha}^{\dagger}=e^{-i\mathbf{k}_{\ell}\cdot\bm{\tau}_{\ell,\alpha}}c_{\ell,\mathbf{k}_{\ell},\alpha}^{\dagger}$.
Using the convention of Eq.~(\ref{eq:alternative_Bloch_convention}),
Eq.~(\ref{eq:wavefunction_real_space_vdW}) would be written as 
\begin{multline}
\psi_{\mathbf{k},n}^{\text{vdW}}(\mathbf{r})=\frac{1}{\sqrt{N_{t}}}\sum_{i,j,\alpha}\tilde{\phi}_{t,\mathbf{k},\alpha}^{n}\left(\mathbf{G}_{b,i}\right)e^{i\left(\mathbf{k}+\mathbf{G}_{b,i}\right)\cdot\mathbf{R}_{t,j}}\times\\
\times w_{t,\alpha}\left(\mathbf{r}-\mathbf{R}_{t,j}-\bm{\tau}_{t,\alpha}\right)+\\
+\frac{1}{\sqrt{N_{b}}}\sum_{i,j,\alpha}\tilde{\phi}_{b,\mathbf{k},\alpha}^{n}\left(\mathbf{G}_{t,i}\right)e^{i\left(\mathbf{k}+\mathbf{G}_{t,i}\right)\cdot\mathbf{R}_{b,j}}\times\\
\times w_{b,\alpha}\left(\mathbf{r}-\mathbf{R}_{b,j}-\bm{\tau}_{b,\alpha}\right),
\end{multline}
where $\tilde{\phi}_{t,\mathbf{k},\alpha}^{n}\left(\mathbf{G}_{b,i}\right)=\phi_{t,\mathbf{k},\alpha}^{n}\left(\mathbf{G}_{b,i}\right)e^{i\left(\mathbf{k}+\mathbf{G}_{b,i}\right)\cdot\bm{\tau}_{t,\alpha}}$
and $\tilde{\phi}_{b,\mathbf{k},\alpha}^{n}\left(\mathbf{G}_{t,i}\right)=\phi_{b,\mathbf{k},\alpha}^{n}\left(\mathbf{G}_{t,i}\right)e^{i\left(\mathbf{k}+\mathbf{G}_{t,i}\right)\cdot\bm{\tau}_{b,\alpha}}$.
With these definitions, $\mathcal{M}_{\mathbf{Q},n}$ would be obtained
from Eq.~(\ref{eq:M_Q}) by replacing $\phi_{\ell,\mathbf{Q}_{\perp},\alpha}^{n}\left(\mathbf{0}\right)\rightarrow\tilde{\phi}_{\ell,\mathbf{Q}_{\perp},\alpha}^{n}\left(\mathbf{0}\right)e^{-i\mathbf{Q}_{\perp}\cdot\bm{\tau}_{\ell,\alpha}}$.
Second, we would like to emphasize that the factors of $\Phi_{m_{\alpha}}\left(\phi_{\hat{\mathbf{Q}}}\right)$,
which are included in Eq\@.~(\ref{eq:M_Q}) via $\mathcal{Y}_{l_{\alpha}}^{m_{\alpha}}\left(\hat{\mathbf{Q}}\right)$,
ensure that the ARPES intensity has the same symmetries under rotations
along the $z$ direction as the bilayer vdW structure .

We would like to point out that the present formalism can also be
applied to model ARPES in other system subject to competing periodicities,
such as systems displaying CDWs \citep{Flicker_2016} or 2D materials
weakly coupled to a substrate.

\begin{figure}
\begin{centering}
\includegraphics[width=8cm]{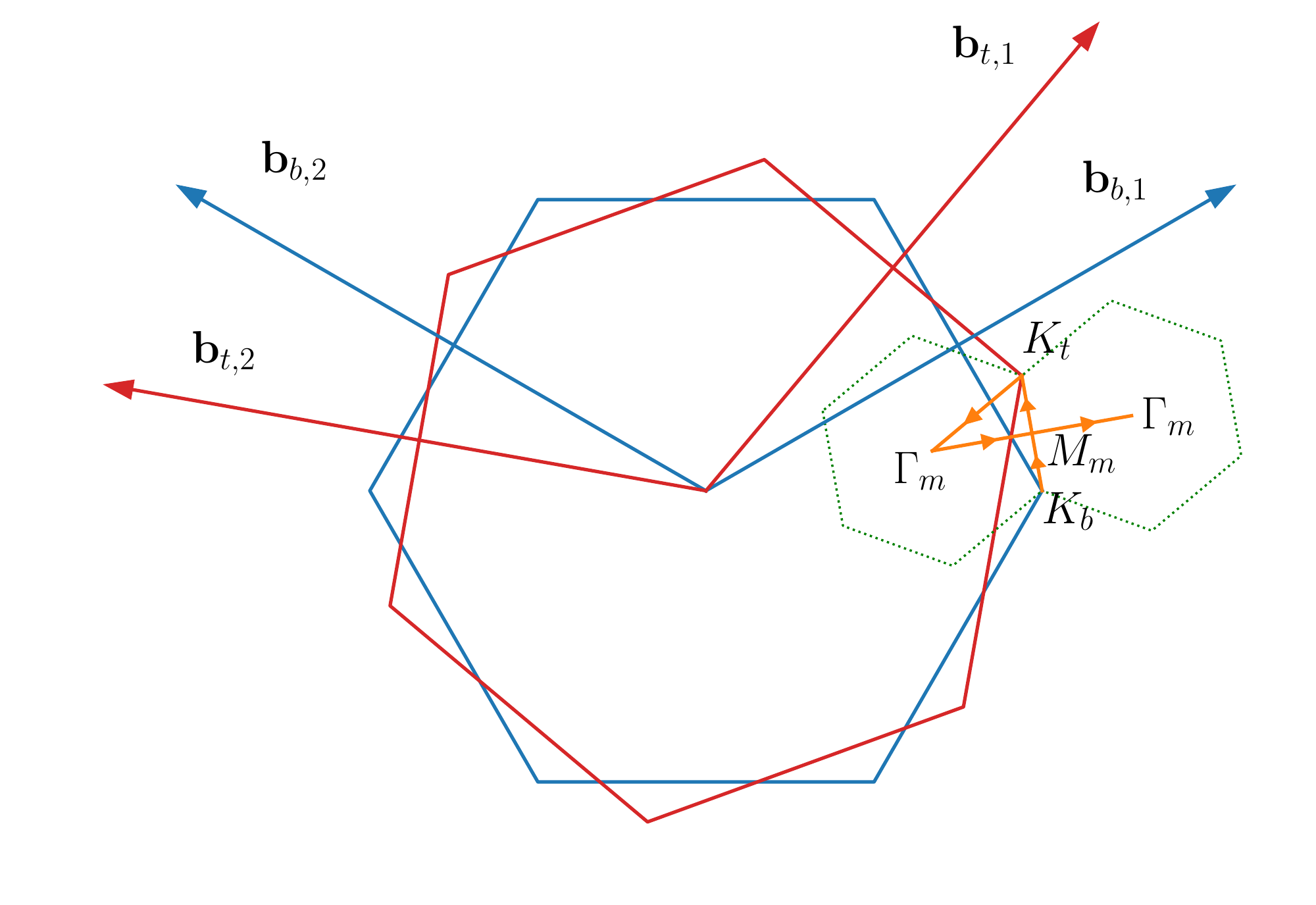} 
\par\end{centering}
\caption{\label{fig:Brillouin-zone}Brillouin zone and basis vectors of reciprocal
lattice for the top (in red) and bottom (in blue) isolated graphene
layers that form a twisted bilayer graphene vdW structure. The dashed
green hexagon shows the Brillouin zone of the Moir{é} superlattice.
The yellow arrows show the path along which the ARPES mapped band
structure of Fig.~\ref{fig:arpes-bands} is computed. }
\end{figure}

\section{Applications\label{sec:Applications}}

We will now apply the general formalism developed in the previous
section to two system: twisted bilayer graphene and twisted bilayer
MoS$_{2}$.

\subsection{ARPES of twisted bilayer graphene\label{subsec:An-application-graphene}}

Graphene has a triangular Bravais lattice, with a unit cell containing
two carbon atoms, A and B, which form a honeycomb structure. We write
the basis vectors for the bottom layer as 
\begin{align}
\mathbf{a}_{b,1} & =a\left(\frac{1}{2},\frac{\sqrt{3}}{2}\right),\label{eq:lattice_a1}\\
\mathbf{a}_{b,2} & =a\left(-\frac{1}{2},\frac{\sqrt{3}}{2}\right),\label{eq:lattice_a2}
\end{align}
where $a\simeq2.46\Ang$ is the graphene lattice parameter, and the
positions of the A and B atoms are given by the sublattice vectors
\begin{align}
\bm{\tau}_{b,A} & =\left(0,0\right),\\
\bm{\tau}_{b,B} & =\frac{a}{\sqrt{3}}\left(0,1\right).
\end{align}
The top layer is rotated with respect to the bottom one by an angle
of $\theta$ such that $\mathbf{a}_{t,i}=R(\theta)\cdot\mathbf{a}_{b,i}$,
for $i=1,2$, and $\bm{\tau}_{t,\alpha}=R(\theta)\cdot\bm{\tau}_{b,\alpha}+d\mathbf{e}_{z}$,
for $\alpha=A,B$, where $R(\theta)$ is the rotation matrix 
\begin{equation}
R(\theta)=\left[\begin{array}{cc}
\cos\theta & -\sin\theta\\
\sin\theta & \cos\theta
\end{array}\right],\label{eq:rotation_matrix}
\end{equation}
and $d\simeq3.35\text{\ensuremath{\Ang}}$ is the separation between
the two layers. The corresponding reciprocal space basis vectors are
given by 
\begin{align}
\mathbf{b}_{b,1} & =\frac{4\pi}{\sqrt{3}a}\left(\frac{\sqrt{3}}{2},\frac{1}{2}\right),\\
\mathbf{b}_{b,2} & =\frac{4\pi}{\sqrt{3}a}\left(-\frac{\sqrt{3}}{2},\frac{1}{2}\right),
\end{align}
for the bottom layer, and by $\mathbf{b}_{t,i}=R(\theta)\cdot\mathbf{b}_{b,i}$
for the top layer. In Fig.~\ref{fig:Brillouin-zone}, we show the
first Brillouin zone of both layers. Also shown is the Brillouin zone
of the moiré superlattice, whose associated reciprocal lattice basis
vectors are given by $\mathbf{b}_{m,i}=\mathbf{b}_{t,i}-\mathbf{b}_{b,i}$
\citep{SanJose_2014a,Amorim_2016_review}. 

\begin{figure}
\begin{centering}
\includegraphics[width=8cm]{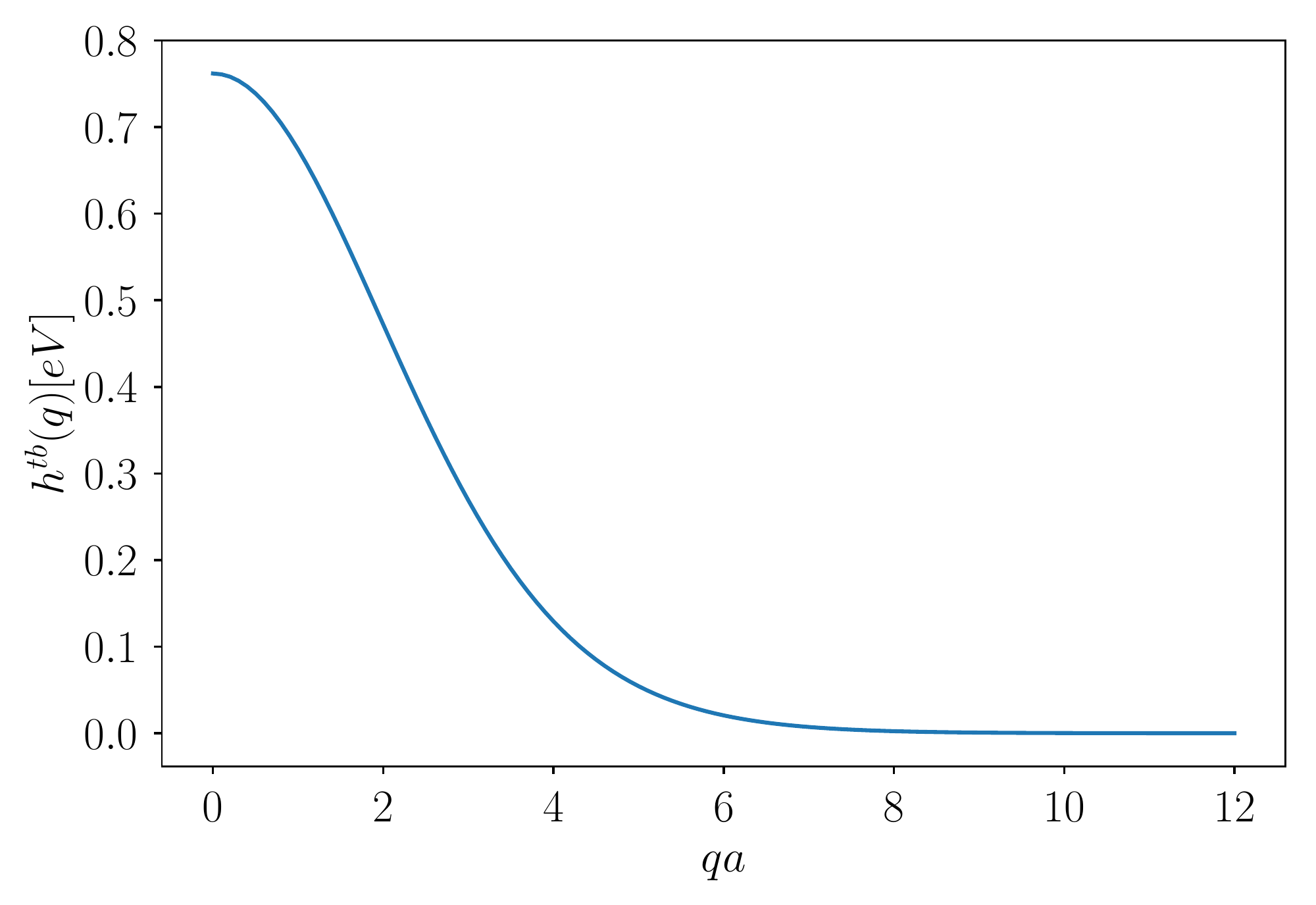} 
\par\end{centering}
\caption{\label{fig:Interlayer-coupling}Plot of the interlayer coupling function
$h^{tb}(\mathbf{q})$ for bilayer graphene, Eq.~(\ref{eq:interlayer_hopping_twistedBG}),
as a function of $qa$, with $a$ the lattice parameter of graphene.
The parameters used are given in the main text.}
\end{figure}

We will describe each individual graphene layer within the nearest-neighbor
tight-binding model for $p_{z}$ orbitals \citep{CastroNeto_2009},
which reads 
\begin{equation}
H_{\ell}=-t\sum_{i}\sum_{j=0}^{2}\left(c_{\ell,\mathbf{R}_{\ell,i},A}^{\dagger}c_{\ell,\mathbf{R}_{\ell,i}+\mathbf{a}_{\ell,j},B}+\text{h.c.}\right),
\end{equation}
where $t\simeq2.7\text{ eV}$ is the nearest-neighbor hopping and
we have written $\mathbf{a}_{\ell,0}=\left(0,0\right)$. For the interlayer
coupling, we have $h_{\alpha\beta}^{t,b}\left(\mathbf{R}_{t,i},\mathbf{R}_{b,j}\right)=h^{tb}\left(\mathbf{R}_{t,i}+\bm{\tau}_{t,\alpha}-\mathbf{R}_{b,j}-\bm{\tau}_{b,\beta}\right)$,
$\alpha,\beta=A,B$, where the function $h^{tb}(\mathbf{r})$ can
be written in terms of Slater-Koster parameters \citep{Slater_Koster_1954}
as 
\begin{equation}
h^{tb}(\mathbf{r})=h^{tb}(\mathbf{r},d)=V_{pp\pi}(R)\frac{r^{2}}{R^{2}}+V_{pp\sigma}(R)\frac{d^{2}}{R^{2}},\label{eq:pz_pz_hopping}
\end{equation}
where $R=\sqrt{r^{2}+d^{2}}$ is the distance between the two atoms,
with $r$ the distance projected in the $x-y$ plane. For the dependence
of the Slater-Koster parameters on $R$, we use the parametrization
of Refs.~\citep{Moon_2013,Koshino_2015}: $V_{pp\pi}(R)=V_{pp\pi}^{0}e^{-\left(R-a/\sqrt{3}\right)/r_{0}}$
and $V_{pp\sigma}(R)=V_{pp\sigma}^{0}e^{-\left(R-d\right)/r_{0}}$,
where $r_{0}\simeq0.184a$, $V_{pp\pi}^{0}=-t\simeq-2.7\text{ eV}$
and $V_{pp\sigma}^{0}\simeq0.48\text{ eV}$. From this we can evaluate
numerically 
\begin{equation}
h_{\alpha\beta}^{tb}(\mathbf{q})=h^{tb}(\mathbf{q})=\frac{1}{A_{\text{c}}}\int d^{2}\mathbf{r}e^{-i\mathbf{q}\cdot\mathbf{r}}h^{tb}\left(\mathbf{r},d\right),\label{eq:interlayer_hopping_twistedBG}
\end{equation}
where $A_{\text{c}}=A_{\text{c},\ell}=\sqrt{3}a^{2}/2$. We plot the
function $h^{tb}(\mathbf{q})$ in Fig.~\ref{fig:Interlayer-coupling}.

Writing the electronic creation and annihilation operators in terms
of Fourier components as in Eq.~(\ref{eq:Bloch-Wannier_states}),
we can construct the Hamiltonian for the graphene bilayer structure
in the form of Eqs.~(\ref{eq:Hamiltonian_vdW_full})-(\ref{eq:Hamiltonian_vdW_top_bottom}),
where 
\begin{equation}
\bm{h}_{\mathbf{k}}^{\ell,\ell}=\left[\begin{array}{cc}
0 & \gamma_{\ell,\mathbf{k}}\\
\gamma_{\ell,\mathbf{k}}^{*} & 0
\end{array}\right],
\end{equation}
is written in the $A,B$ sublattice basis, $\gamma_{\ell,\mathbf{k}}=-t\sum_{i=1}^{3}e^{i\mathbf{k}\cdot\bm{\delta}_{\ell,i}}$,
with $\bm{\delta}_{\ell,i}=R\left(2\pi\left(i-1\right)/3\right)\cdot\bm{\tau}_{\ell,B}$
the nearest-neighbor vectors in each layer, and the interlayer coupling
matrices are given by 
\begin{multline}
\bm{h}_{\mathbf{k},\mathbf{G}_{b},\mathbf{G}_{t}}^{t,b}=\left[\begin{array}{cc}
e^{i\mathbf{G}_{t}\cdot\bm{\tau}_{t,A}} & 0\\
0 & e^{i\mathbf{G}_{t}\cdot\bm{\tau}_{t,B}}
\end{array}\right]\cdot\\
\cdot\left[\begin{array}{cc}
h^{tb}(\mathbf{k}+\mathbf{G}_{b}+\mathbf{G}_{t}) & h^{tb}(\mathbf{k}+\mathbf{G}_{b}+\mathbf{G}_{t})\\
h^{tb}(\mathbf{k}+\mathbf{G}_{b}+\mathbf{G}_{t}) & h^{tb}(\mathbf{k}+\mathbf{G}_{b}+\mathbf{G}_{t})
\end{array}\right]\cdot\\
\cdot\left[\begin{array}{cc}
e^{-i\mathbf{G}_{b}\cdot\bm{\tau}_{b,A}} & 0\\
0 & e^{-i\mathbf{G}_{b}\cdot\bm{\tau}_{b,B}}
\end{array}\right],
\end{multline}
with $\bm{h}_{\mathbf{k},\mathbf{G}_{t},\mathbf{G}_{b}}^{b,t}=\left(\bm{h}_{\mathbf{k},\mathbf{G}_{b},\mathbf{G}_{t}}^{t,b}\right)^{\dagger}$.
By truncating the Hamiltonian to the $N_{G}$ shortest $\mathbf{G}_{t}$
and $\mathbf{G}_{b}$ vectors, $\bm{H}_{\mathbf{k}}\left(\left\{ \mathbf{G}_{b}\right\} ,\left\{ \mathbf{G}_{t}\right\} \right)$
becomes a $4N_{G}\times4N_{G}$ matrix, from which we can obtain an
approximation to the energies and eigenstates of the twisted bilayer
graphene structure. The obtained model is a tight-binding extension
of the continuous models of Refs.~\citep{dosSantos_2007,Bistritzer_2011,dosSantos_2012,Moon_2013}.
From the eigenstates, we can obtain the ARPES intensity using Eqs.~(\ref{eq:ARPES_intensity_final})
and (\ref{eq:M_Q}). Due to the fact that the function $h^{tb}(\mathbf{q})$
decays rapidly for $\left|\mathbf{q}\right|a\gg1$, the obtained ARPES
signal converges rapidly with only a few $\mathbf{G}_{t}$ and $\mathbf{G}_{b}$
vectors. Notice that due to the fact that the model we use only involves
$p_{z}\equiv\mathcal{Y}_{1}^{0}$ orbitals, we have that $\Phi_{0}\left(\phi_{\hat{\mathbf{Q}}}\right)=1/\sqrt{2}$
in Eq.~(\ref{eq:M_Q}) for $\mathcal{M}_{\mathbf{Q},n}$.

\begin{figure}
\begin{centering}
\includegraphics[width=8cm]{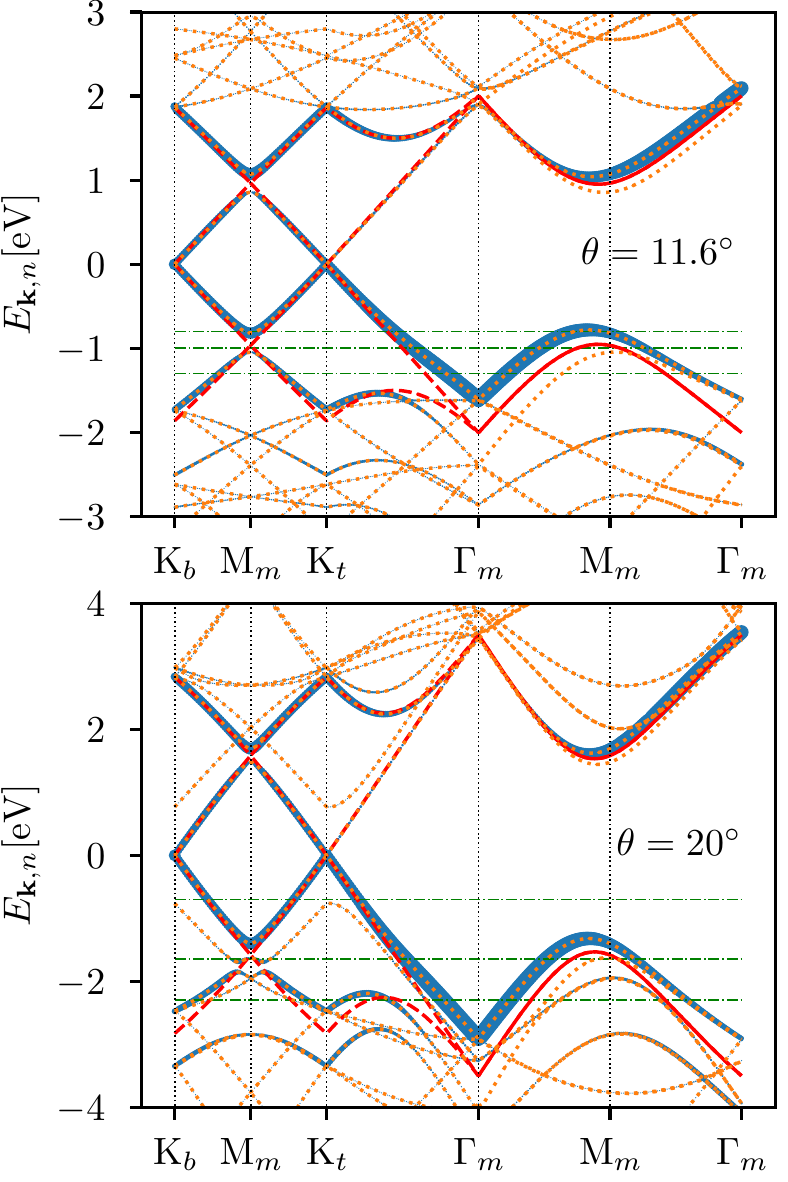} 
\par\end{centering}
\caption{\label{fig:arpes-bands}Computed ARPES bands for twisted bilayer graphene
for two different twist angles: $\theta=11.6^{\circ}$ (top) and $\theta=20^{\circ}$
(bottom). The yellow dotted lines show the bands $E_{\mathbf{k},n}$,
eigenvalues of $\bm{H}_{\mathbf{k}}\left(\left\{ \mathbf{G}_{b}\right\} ,\left\{ \mathbf{G}_{t}\right\} \right)$,
for $\mathbf{k}$ along the path $\text{K}_{b}\rightarrow\text{M}_{m}\rightarrow\text{K}_{t}\rightarrow\Gamma_{m}\rightarrow\text{M}_{m}\rightarrow\Gamma_{m}$
represented by the yellow arrows in Fig.~\ref{fig:Brillouin-zone}.
The blue tick bands represent $E_{\mathbf{k},n}$ weighted by the
ARPES matrix element value, with the thickness corresponding to $\left|\mathcal{M}_{\mathbf{k},0,n}\right|^{2}$.
The dashed red lines shown the dispersion relation for the decoupled
graphene layers. It was assumed $Q_{z}=0$. The horizontal dot-dashed
lines mark the energies $-1.3$, $-1.0$ and $-0.8\text{ eV}$ (for
$\theta=11.6^{\circ}$) and $-2.3$, $-1.65$ and $-0.7\text{ eV}$
(for $\theta=20^{\circ}$) at which the ARPES constant energy maps
of Fig.~\ref{fig:arpes-constant-energy-11.6o} are computed. In both
plots we used a number of reciprocal lattice vectors $N_{G}=7$. It
was assumed $Q_{z}=0$.}
\end{figure}

\begin{figure}
\begin{centering}
\includegraphics[width=8cm]{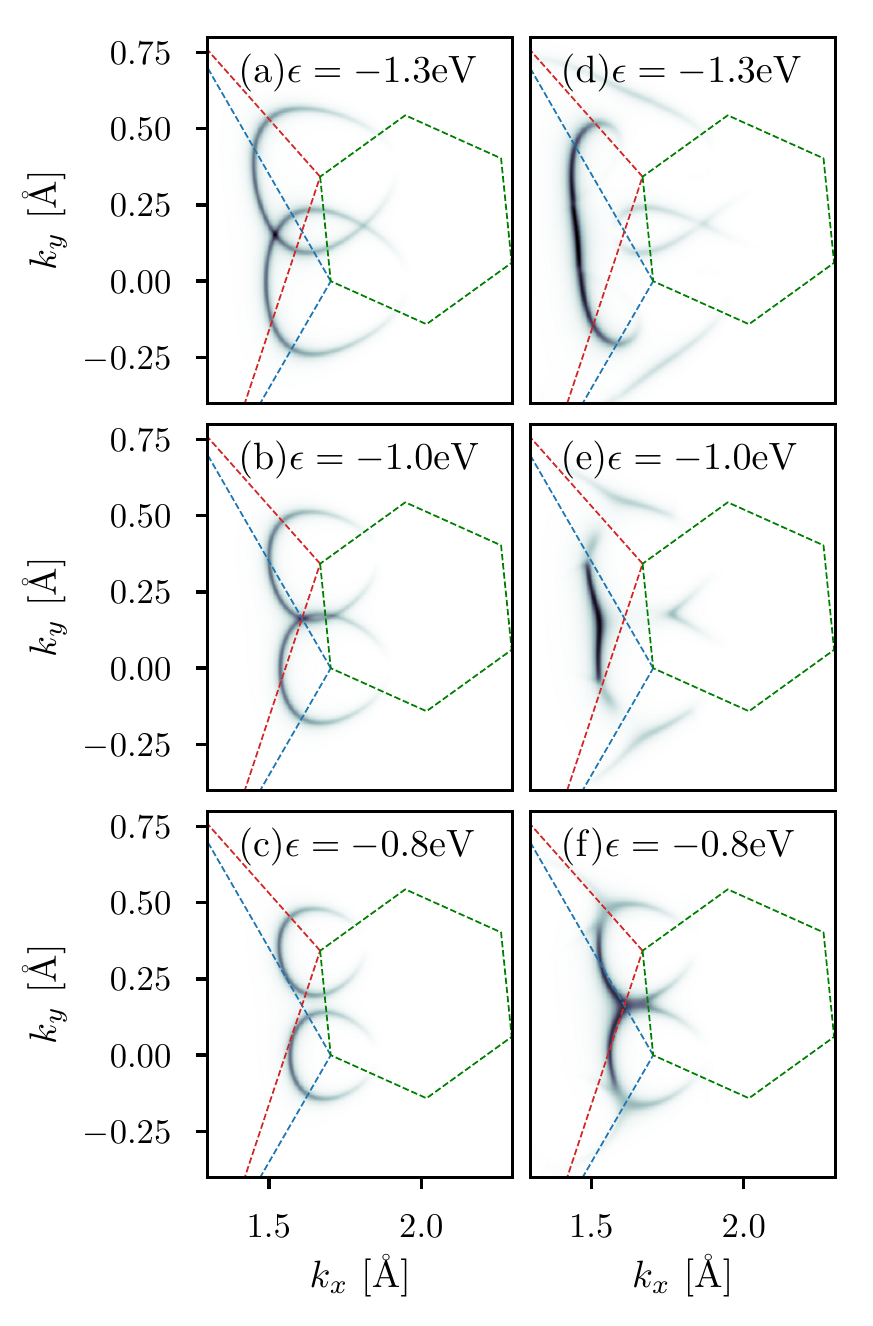} 
\par\end{centering}
\caption{\label{fig:arpes-constant-energy-11.6o}Computed constant energy ARPES
map for twisted bilayer graphene for a twist angle of $\theta=11.6^{\circ}$
for the energies of $\epsilon=-1.3$, $-1.0$ and $-0.8$ eV measured
from the Dirac point. (a)--(c) show the constant energy ARPES maps
for two decoupled graphene layers, while (d)--(f) show the ARPES
maps taking into account coupling between the two layers. The dashed
red and blue lines show the limits of the Brillouin zone of the top
and bottom layers (respectively), while the green dashed hexagon represents
the moiré Brillouin zone. A broadening of $\eta=0.05$ eV was used.
The Hamiltonian was truncated with $N_{G}=7$. It was assumed $Q_{z}=0$.}
\end{figure}

\begin{figure}
\begin{centering}
\includegraphics[width=8cm]{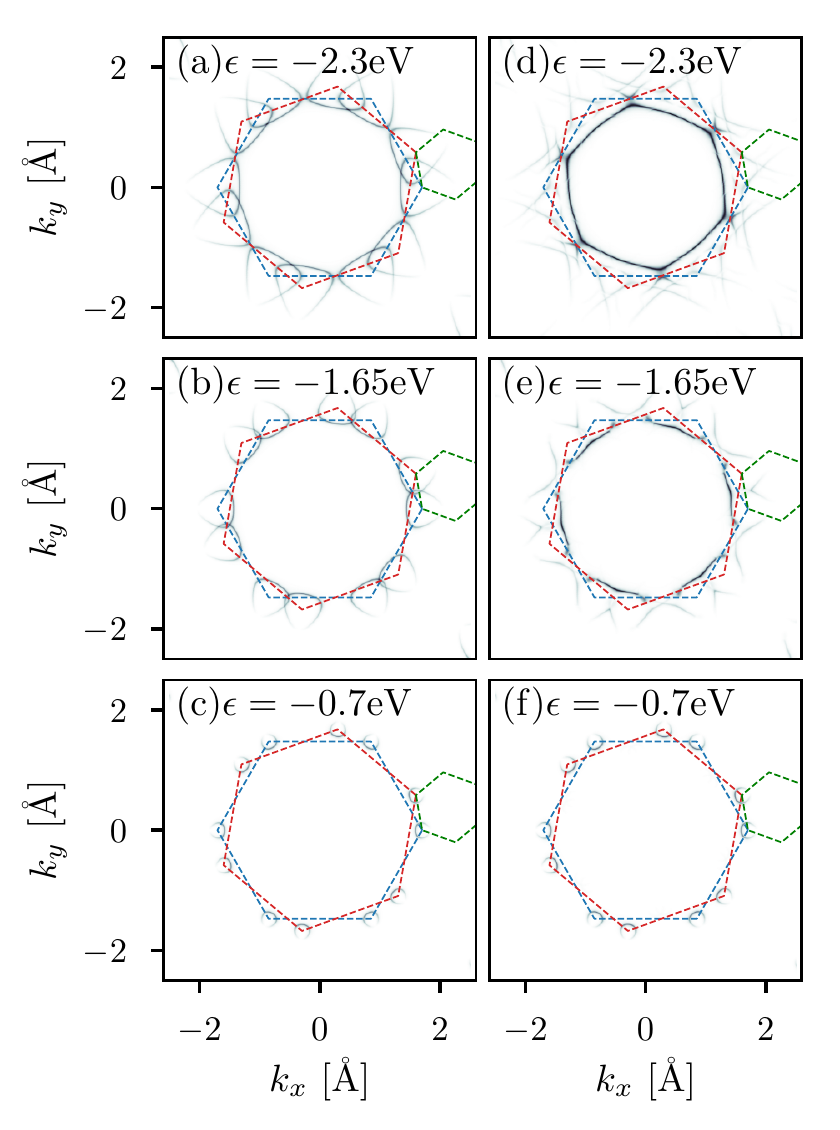} 
\par\end{centering}
\caption{\label{fig:arpes-constant-energy-20o}Computed constant energy ARPES
map for twisted bilayer graphene for a twist angle of $\theta=20^{\circ}$
for the energies of $\epsilon=-2.3$, $-1.65$ and $-0.7$ eV measured
from the Dirac point. (a)--(c) show the constant energy ARPES maps
for two decoupled graphene layers, while (d)--(f) show the ARPES
maps taking into account coupling between the two layers. The dashed
red, blue and green lines represent the same as in Fig.~\ref{fig:arpes-constant-energy-11.6o}.
A broadening of $\eta=0.05$ eV was used. The Hamiltonian was truncated
with $N_{G}=7$. It was assumed $Q_{z}=0$.}
\end{figure}

In Fig.~\ref{fig:arpes-bands} we show the computed ARPES bands along
the path indicated in Fig.~\ref{fig:Brillouin-zone}, for two different
angles: $\theta=11.6^{\circ}$, for which ARPES measurements where
performed in Ref.~\citep{Ohta_2012}, and for $\theta=20^{\circ}$.
The thickness of the bands is proportional to the value of $\left|\mathcal{M}_{\mathbf{Q}_{\perp},0,n}\right|^{2}$,
where for simplicity we assumed that there is no transferred momentum
along the $z$ direction, $Q_{z}=0$, and neglected the effect of
$\tilde{R}_{\ell,\alpha}\left(Q\right)$ on $\mathcal{M}_{\mathbf{Q}_{\perp},0,n}$.
The bands where computed using a truncated Hamiltonian with $N_{G}=7$
{[}including $\mathbf{G}_{b/t}=\left(0,0\right)${]}, which was found
to be sufficient to obtain converged results. Increasing $N_{G}$
virtually does not change the thick bands in a visible away, although
more $E_{\mathbf{k},n}$ bands do appear, which, nevertheless, have
negligible ARPES weight. As anticipated the ARPES weighted bands mostly
follow the bands of the decoupled system. Constant energy ARPES maps
were also computed for the energies signaled by the dot-dashed horizontal
lines in Fig.~\ref{fig:arpes-bands}. The computed constant energy
maps are shown in Figs.~\ref{fig:arpes-constant-energy-11.6o} (for
$\theta=11.6^{\circ}$) and \ref{fig:arpes-constant-energy-20o} (for
$\theta=20^{\circ}$). For comparison, we also show the constant energy
maps for the decoupled bilayer structures. The absence of part of
the constant energy surface, due to the ARPES matrix elements, in
both the coupled and decoupled cases is clear. A significant reconstruction
of the constant energy ARPES maps is observed for energies at which
the band structures of the isolated graphene layers intersect.

Good qualitative agreement between our results for the twist angle
of $\theta=11.6^{\circ}$ and the experimental data of Ref.~\citep{Ohta_2012}
is observed, especially taking into account the simplicity of our
model. There is however an observed discrepancy between our results
and the experimental data: in Ref.~\citep{Ohta_2012} two bands are
observed with energy $\simeq-1\text{eV}$ along the line that bisects
the angle between the Dirac points of the two layers (Fig.~2(d) of
Ref.~\citep{Ohta_2012}), which is shown as the path $\Gamma_{m}\rightarrow\text{M}_{m}\rightarrow\Gamma_{m}$
in Fig.~\ref{fig:Brillouin-zone}, while in our model there is only
one band with ARPES weight around that energy. The absence of one
of the ARPES bands in our results is easily understandable: the two
bands are formed by bands of the top and bottom layers which are degenerate
for the decoupled system along the path $\Gamma_{m}\rightarrow\text{M}_{m}\rightarrow\Gamma_{m}$
(see the merging of two bands of the decoupled system along that line
in Fig.~\ref{fig:arpes-bands}) corresponding, therefore, to bonding
and anti-bonding states. For the anti-bonding state we can check numerically
that $\phi_{b,\mathbf{k},A}^{\text{anti}}\left(\mathbf{0}\right)=-\phi_{t,\mathbf{k},B}^{\text{anti}}\left(\mathbf{0}\right)$
and $\phi_{b,\mathbf{k},B}^{\text{anti}}\left(\mathbf{0}\right)=-\phi_{t,\mathbf{k},A}^{\text{anti}}\left(\mathbf{0}\right)$.
\footnote{The exchange of sublattices is due to the fact that the path $\Gamma_{m}\rightarrow\text{M}_{m}\rightarrow\Gamma_{m}$
for the top layer is equivalent to the path obtained by exchanging
$y\rightarrow-y$ for the bottom layer. With the chosen sublattice
basis, the operation $y\rightarrow-y$ leads to the change $A\leftrightarrow B$.} For $Q_{z}=0$, we then obtain $\mathcal{M}_{\mathbf{k},0,\text{anti}}\propto\phi_{t,\mathbf{k},A}^{\text{anti}}\left(\mathbf{0}\right)+\phi_{t,\mathbf{k},B}^{\text{anti}}\left(\mathbf{0}\right)+\phi_{b,\mathbf{k},A}^{\text{anti}}\left(\mathbf{0}\right)+\phi_{b,\mathbf{k},B}^{\text{anti}}\left(\mathbf{0}\right)=0$,
and therefore this band would not be visible in ARPES. There are two
possible explanations for the fact that the anti-bonding band is visible
experimentally in ARPES: (i) a possible energy imbalance between the
two layers, (ii) effects of finite transferred momentum along the
out-of-plane direction, $Q_{z}$. As a matter of fact, a shift in
energy between the Dirac cones of the two layers of $\Delta E=0.05$
eV is reported in Ref.~\citep{Ohta_2012}. We have checked that this
shift in energy indeed leads to $\mathcal{M}_{\mathbf{k},0,\text{anti}}\neq0$,
but the value is too small to lead to a significant visibility of
the anti-bonding band. The remaining possibility is finite $Q_{z}$
effects. For finite $Q_{z}$, we would obtain (assuming the Dirac
points of both layers are aligned) 

\begin{multline}
\mathcal{M}_{\mathbf{k},Q_{z},\text{anti}}\propto e^{-iQ_{z}d}\left(\phi_{t,\mathbf{k},A}^{\text{anti}}\left(\mathbf{0}\right)+\phi_{t,\mathbf{k},B}^{\text{anti}}\left(\mathbf{0}\right)\right)\\
+\phi_{b,\mathbf{k},A}^{\text{anti}}\left(\mathbf{0}\right)+\phi_{b,\mathbf{k},B}^{\text{anti}}\left(\mathbf{0}\right)\\
=e^{-iQ_{z}d/2}2\sin\left(Q_{z}d/2\right)\\
\times\left(\phi_{b,\mathbf{k},B}^{\text{anti}}\left(\mathbf{0}\right)+\phi_{b,\mathbf{k},A}^{\text{anti}}\left(\mathbf{0}\right)\right),
\end{multline}
while for the bonding state, for which $\phi_{b,\mathbf{k},A}^{\text{bond}}\left(\mathbf{0}\right)=\phi_{t,\mathbf{k},B}^{\text{bond}}\left(\mathbf{0}\right)$
and $\phi_{b,\mathbf{k},B}^{\text{bond}}\left(\mathbf{0}\right)=\phi_{t,\mathbf{k},A}^{\text{bond}}\left(\mathbf{0}\right)$,
we would obtain 
\begin{multline}
\mathcal{M}_{\mathbf{k},Q_{z},\text{bond}}\propto e^{-iQ_{z}d/2}2\cos\left(Q_{z}d/2\right)\\
\times\left(\phi_{b,\mathbf{k},A}^{\text{anti}}\left(\mathbf{0}\right)+\phi_{b,\mathbf{k},B}^{\text{anti}}\left(\mathbf{0}\right)\right).
\end{multline}
Therefore, both bands can have similar visibility in ARPES if $Q_{z}d/2\simeq\left(1+2n\right)\pi/4$,
$n\in\mathbb{N}$. Besides this effect, we also expect that a better
agreement with the experimental data would be possible, provided a
more accurate modeling of the band structure of the individual layers
was employed.

\subsection{ARPES of twisted bilayer MoS$_{2}$\label{subsec:An-application-MoS2}}

Similarly to graphene, monolayer MoS$_{2}$ has a honeycomb structure,
with the A sites occupied by Mo atoms and the B sites occupied by
two S atoms, top and bottom, which lie at planes above and bellow
the Mo plane. We write the Bravais basis vectors for MoS$_{2}$ in
the same way as for graphene, Eqs.~\eqref{eq:lattice_a1} and \eqref{eq:lattice_a2},
with a lattice constant $a\simeq3.16\Ang$\citep{Cappelluti_2013}.
For an unrotated layer, the Mo and S atoms occupy the approximate
positions inside the unit cell 
\begin{align}
\bm{\tau}_{\text{Mo}} & =\left(0,0\right),\\
\bm{\tau}_{\text{S}^{\text{top}/\text{bot}}} & =a\left(\frac{1}{\sqrt{3}},\pm\frac{1}{2}\right).
\end{align}
For the separation between the two MoS$_{2}$ layers (between the
Mo planes) we use the value for bulk MoS$_{2}$ $c^{\prime}\simeq6.14\Ang$,
which corresponds to a separation between nearest S planes of $d\simeq2.98\Ang$.
We will describe the electronic properties of the individual MoS$_{2}$
layers, using the 11 band tight-binding Hamiltonian of Ref.~\citep{Cappelluti_2013},
with the parametrization of Ref.~\citep{Roldan_2014}, which involves
the $d_{x^{2}-y^{2}}$, $d_{xy}$ $d_{xz}$, $d_{yz}$, $d_{z^{2}}$
Mo orbitals (corresponding to the real spherical harmonics $\mathcal{Y}_{2}^{2}$,
$\mathcal{Y}_{2}^{-2}$ $\mathcal{Y}_{2}^{1}$, $\mathcal{Y}_{2}^{-1}$,
$\mathcal{Y}_{2}^{0}$) and the $p_{x}$, $p_{y}$, $p_{z}$ S orbitals
(corresponding to the real spherical harmonics $\mathcal{Y}_{1}^{1}$,
$\mathcal{Y}_{1}^{-1}$, $\mathcal{Y}_{1}^{0}$). The Hamiltonian
for an unrotated layer is given by
\begin{equation}
H=\sum_{n=0}^{6}\bm{c}_{\mathbf{R}+\mathbf{a}_{n}}^{\dagger}\cdot\bm{h}_{n}\cdot\bm{c}_{\mathbf{R}},
\end{equation}
where
\begin{multline}
\bm{c}_{\mathbf{R}}^{\dagger}=\left(c_{\mathbf{R},d_{x^{2}-y^{2}}}^{\dagger},\,c_{\mathbf{R},d_{xy}}^{\dagger},\,c_{\mathbf{R},d_{xz}}^{\dagger},\,c_{\mathbf{R},d_{yz}}^{\dagger},\,c_{\mathbf{R},d_{z^{2}}}^{\dagger},\,\right.\\
\left.c_{\mathbf{R},p_{x}^{\text{top}}}^{\dagger},\,c_{\mathbf{R},p_{y}^{\text{top}}}^{\dagger},\,c_{\mathbf{R},p_{z}^{\text{top}}}^{\dagger},\,c_{\mathbf{R},p_{x}^{\text{bot}}}^{\dagger},\,c_{\mathbf{R},p_{y}^{\text{bot}}}^{\dagger},\,c_{\mathbf{R},p_{z}^{\text{bot}}}^{\dagger}\right),\label{eq:MoS2_basis}
\end{multline}
is a vector of creation operators, $\bm{h}_{n}$ are hoping matrices
and $\mathbf{a}_{n}$ are given by $\mathbf{a}_{n}=R\left(\pi(n-1)/3\right)\cdot\mathbf{a}_{1}$
for $n=1,...,6$ and $\mathbf{a}_{0}=\left(0,0\right)$. Writing the
electronic creation and annihilation operators in terms of Fourier
components as in Eq.~(\ref{eq:Bloch-Wannier_states}) we obtain the
following Hamiltonian matrix in $\mathbf{k}$-space, $\bm{h}(\mathbf{k})$,
with entries $h_{\alpha\beta}\left(\mathbf{k}\right)=\sum_{n=0}^{6}\left[\bm{h}_{n}\right]_{\alpha\beta}e^{-i\mathbf{k}\cdot\left(\mathbf{a}_{n}+\bm{\tau}_{\alpha}-\mathbf{\bm{\tau}_{\beta}}\right)}$,
with $\alpha,\beta=d_{x^{2}-y^{2}},...,p_{z}^{\text{bot}}$. As in
the case of graphene, we will keep the bottom layer fixed, while rotating
the top layer by an angle $\theta$. When describing a MoS$_{2}$
layer rotated by an angle $\theta$, it is important to take into
account that under the rotation the orbitals will transform in a non-trivial
way (in graphene this does not happen as $p_{z}$ orbitals are invariance
under rotations around the $z$ axis). We chose to represent the Hamiltonian
of both layers in terms of orbitals defined with respect to the same
common reference frame, which we choose to be the unrotated frame.
It is also with respect to this common reference frame that the plane
wave expansion Eq.~\eqref{eq:plane_wave_expansion} is written. Taking
this into account, we can write the Hamiltonian matrix for each layer,
in the orbital basis defined with respect to the common reference
frame, as
\begin{align}
\bm{h}_{\mathbf{k}}^{b,b} & =\bm{h}\left(\mathbf{k}\right),\label{eq:hamiltonian_MoS2_unrotated}\\
\bm{h}_{\mathbf{k}}^{t,t} & =\bm{\mathcal{R}}(\theta)\cdot\bm{h}\left(R(-\theta)\cdot\mathbf{k}\right)\cdot\mathcal{\bm{R}}(-\theta),\label{eq:hamiltonian_MoS2_rotated}
\end{align}
where the matrix $\bm{\mathcal{R}}(\theta)$ rotates the orbitals
in Eq.~\eqref{eq:MoS2_basis} along the $z$ axis, and has the block
diagonal form
\begin{equation}
\bm{\mathcal{R}}(\theta)=\left[\begin{array}{ccccccc}
R(2\theta)\\
 & R(\theta)\\
 &  & 1_{1\times1}\\
 &  &  & R(\theta)\\
 &  &  &  & 1_{1\times1}\\
 &  &  &  &  & R(\theta)\\
 &  &  &  &  &  & 1_{1\times1}
\end{array}\right],
\end{equation}
with $R(\theta)$ the rotation matrix Eq.~\eqref{eq:rotation_matrix}.
For the interlayer coupling, we assume that this is dominated by the
hopping between the $\text{S}^{\text{top}}$ $p$ orbitals of the
bottom MoS$_{2}$ layer and the $\text{S}^{\text{bot}}$ $p$ orbitals
of the MoS$_{2}$ layer. We write the interlayer hoppings in terms
of Slater-Koster parameters as
\begin{multline}
h_{p_{i}^{\text{bot}},p_{j}^{\text{top}}}^{tb}(\mathbf{r})=h_{p_{i}^{\text{bot}},p_{j}^{\text{top}}}^{tb}(\mathbf{r},d)\\
=V_{pp\sigma}\left(R\right)\frac{R^{i}R^{j}}{R^{2}}+V_{pp\pi}\left(R\right)\left(\delta_{ij}-\frac{R^{i}R^{j}}{R^{2}}\right),\\
\,i,j=x,y,z\label{eq:pp_interlayer}
\end{multline}
where $\mathbf{R}=\left(x,y,d\right)$ is the separation between the
S atoms. As before, we assume a dependence of the Slater-Koster parameters
on the distance of the form $V_{pp\sigma}\left(R\right)=V_{pp\sigma}^{0}e^{-\beta\left(R/d_{SS}-1\right)}$
and $V_{pp\pi}\left(R\right)=V_{pp\pi}^{0}e^{-\beta\left(R/d_{SS}-1\right)}$.
We use $V_{pp\sigma}^{0}\simeq-0.774$ eV and $V_{pp\pi}^{0}\simeq0.123$
eV, which are the interlayer hoppings used in Ref.~\citep{Cappelluti_2013},
with $d_{SS}\simeq3.49\Ang$ the interlayer nearest-neighbor separation
between S atoms in bulk MoS$_{2}$. In the absence of \textit{ab initio}
calculations, we assume the value $\beta\simeq3$, in accordance with
Harrison's argument \citep{Harrison_book} and as previously used
to model strained MoS$_{2}$ \citep{Castellanos2103}. The Fourier
transform of $h_{p_{i}^{\text{bot}},p_{j}^{\text{top}}}^{tb}(\mathbf{r},d)$
can be written as
\begin{multline}
h_{p_{i}^{\text{bot}},p_{j}^{\text{top}}}^{tb}(\mathbf{q})=\\
=\left[\begin{array}{ccc}
V_{1}(q)-\frac{q_{x}^{2}}{q^{2}}V_{2}(q) & -\frac{q_{x}q_{y}}{q^{2}}V_{2}(q) & -i\frac{q_{x}}{q}V_{z}(q)\\
-\frac{q_{x}q_{y}}{q^{2}}V_{2}(q) & V_{1}(q)-\frac{q_{y}^{2}}{q^{2}}V_{2}(q) & -i\frac{q_{y}}{q}V_{z}(q)\\
-i\frac{q_{x}}{q}V_{z}(q) & -i\frac{q_{y}}{q}V_{z}(q) & V_{zz}(q)
\end{array}\right],\label{eq:interlayer_MoS2}
\end{multline}
where the functions $V_{zz}(q)$, $V_{z}(q)$, $V_{1}(q)$ and $V_{2}(q)$
are shown in Fig.~\ref{fig:Interlayer-coupling-MoS2} and the explicit
expressions used to evaluate them are provided in Appendix~\ref{appx:interlayer_coupling}.
We point out that the general expression for $V_{zz}(q)$ is the same
as for graphene, Eq.~\eqref{eq:interlayer_hopping_twistedBG}. As
for graphene, we can see that the Fourier components of the interlayer
hoppings decay fast. Having evaluated $h_{p_{i}^{\text{bot}},p_{j}^{\text{top}}}^{tb}(\mathbf{q})$
we can construct the matrices $\bm{h}_{\mathbf{k},\mathbf{G}_{b},\mathbf{G}_{t}}^{t,b}$
and $\bm{h}_{\mathbf{k},\mathbf{G}_{t},\mathbf{G}_{b}}^{b,t}$, from
which, together with Eqs.~\eqref{eq:hamiltonian_MoS2_unrotated}
and \eqref{eq:hamiltonian_MoS2_rotated}, we can build the Hamiltonian
matrix for the twisted bilayer according to Eqs.~\eqref{eq:Hamiltonian_vdW_full}-\eqref{eq:Hamiltonian_vdW_top_bottom}.
Keeping the $N_{G}$ shortest $\mathbf{G}_{t}$ and $\mathbf{G}_{b}$
vectors, $\bm{H}_{\mathbf{k}}\left(\left\{ \mathbf{G}_{b}\right\} ,\left\{ \mathbf{G}_{t}\right\} \right)$
becomes an $11N_{G}\times11N_{G}$ matrix, from which the energies
and eigenstates of twisted bilayer MoS$_{2}$ can be evaluated and
then used to model the ARPES intensity. As already pointed out, the
tight-binding model for MoS$_{2}$ involves orbitals that do not transform
trivially under rotations around the $z$ axis. Therefore, it is essential
to keep the factors $\Phi_{m_{\alpha}}\left(\phi_{\hat{\mathbf{Q}}}\right)$,
Eq.~\ref{eq:angular_inplane}, in the ARPES matrix element $\mathcal{M}_{\mathbf{Q},n}$
in order to obtain an ARPES signal that respects the three-fold rotational
invariance of the twisted bilayer MoS$_{2}$ structure.

\begin{figure}
\begin{centering}
\includegraphics[width=8cm]{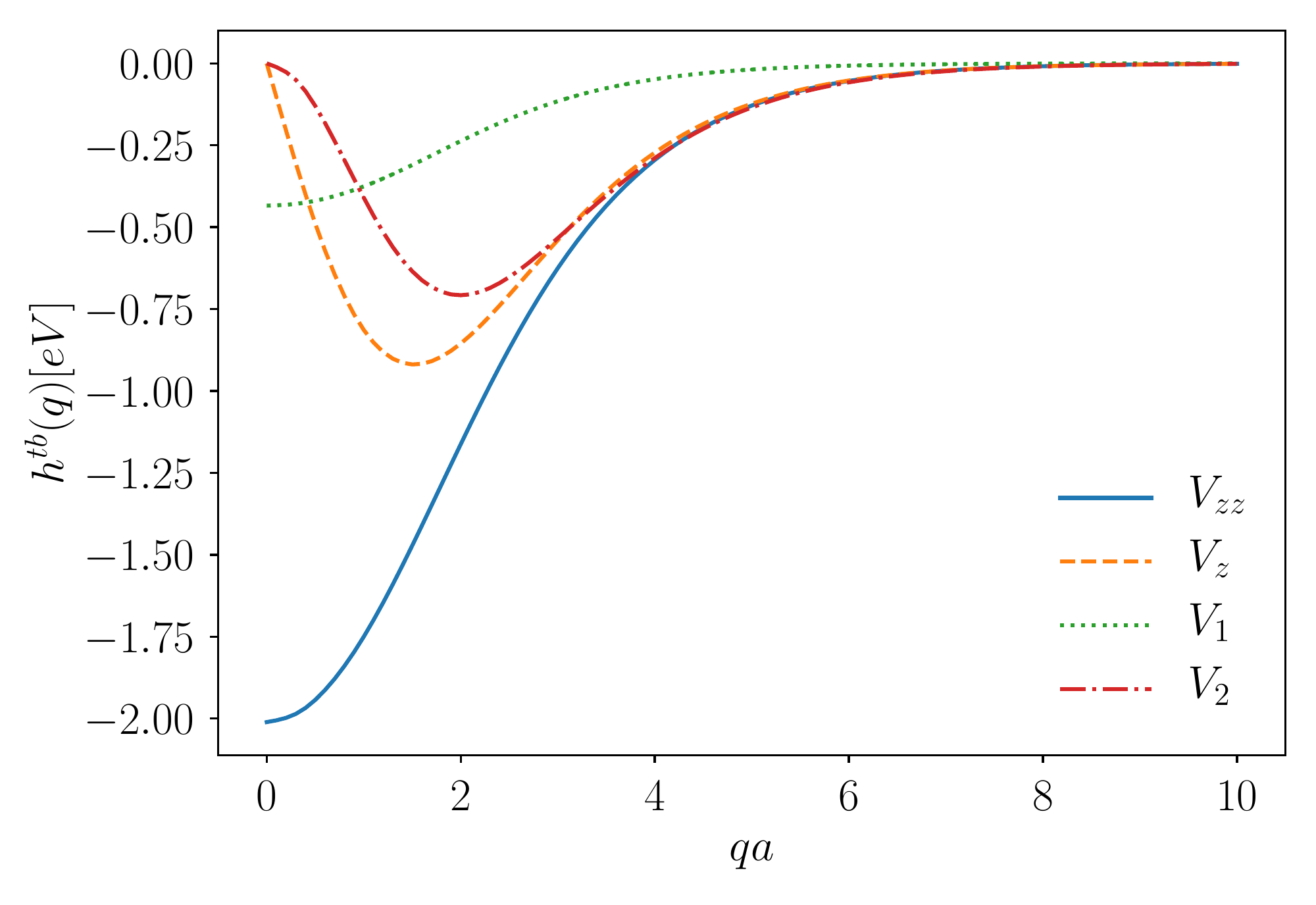} 
\par\end{centering}
\caption{\label{fig:Interlayer-coupling-MoS2}Plot of the interlayer coupling
functions $V_{zz}(q)$, $V_{z}(q)$, $V_{1}(q)$ and $V_{2}(q)$ for
bilayer MoS$_{2}$, Eq.~\eqref{eq:interlayer_MoS2}, as a function
of $qa$, with $a$ the lattice parameter of MoS$_{2}$. The parameters
used are given in the main text.}
\end{figure}

\begin{figure}
\begin{centering}
\includegraphics[width=8cm]{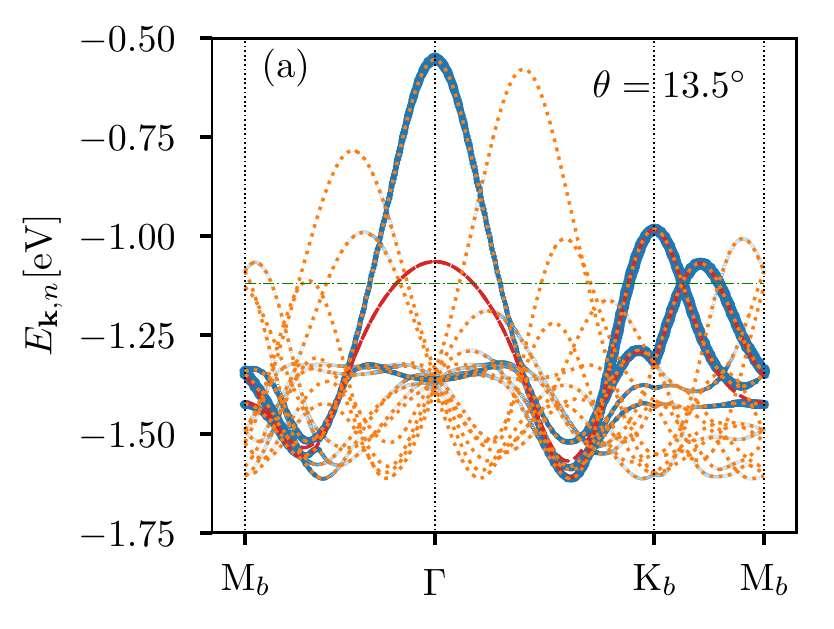}
\par\end{centering}
\begin{centering}
\includegraphics[width=8cm]{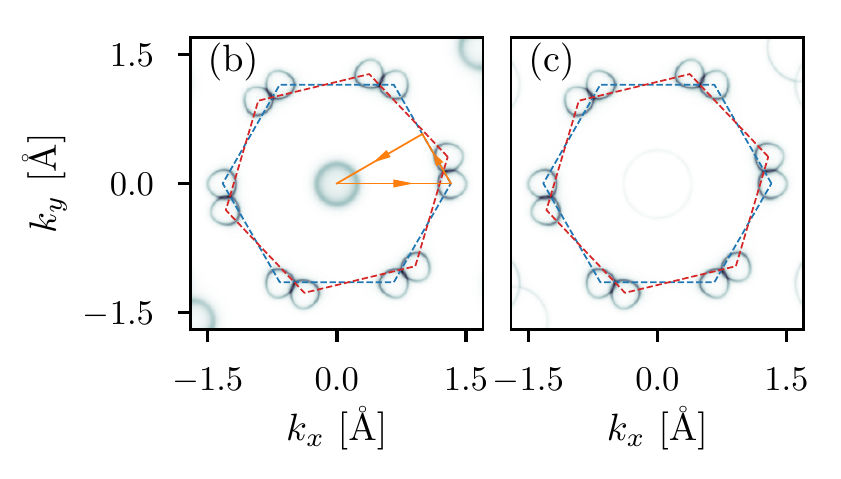}
\par\end{centering}
\caption{\label{fig:arpes-tBMoS2}Computed ARPES bands, (a), and constant energy
map, (b) and (c), for twisted bilayer MoS$_{2}$ for a twist angle
of $\theta=13.5^{\circ}$. In (a), the bands are computed along the
path $\text{M}_{b}\rightarrow\Gamma\rightarrow\text{K}_{b}\rightarrow\text{M}_{b}$
{[}shown in (b){]}. The yellow dotted lines show the band structure,
$E_{\mathbf{k},n}$, while the blue tick bands represent $E_{\mathbf{k},n}$
weighted by the ARPES matrix element value, with the thickness corresponding
to $\left|\mathcal{M}_{\mathbf{k},0,n}\right|^{2}$. The dashed red
lines show the band structure for the decoupled MoS$_{2}$ layers.
The horizontal dot-dashed green line marks the energy $\epsilon=-1.12$
eV, at which the constant energy maps (b) and (c) are computed. (b)
and (c) are, respectively, the constant energy maps for decoupled
and coupled twisted bilayer MoS$_{2}$. A broadening of $\eta=0.02$
eV was used. The Hamiltonian for the coupled bilayer structure was
truncated with $N_{G}=7$. In all plots it was assumed $Q_{z}=0$.}
\end{figure}

In Fig.~\ref{fig:arpes-tBMoS2} we show the computed ARPES bands
and constant energy maps for a twisted bilayer MoS$_{2}$ with a twist
angle of $\theta=13.5^{\circ}$, for which ARPES measurements have
been performed \citep{Yeh_2016}. The calculations where performed
truncating the Hamiltonian matrix with $N_{G}=7$. As can be seen
in Fig.~\ref{fig:arpes-tBMoS2}(a), by comparing with the bands of
the decoupled layers, the interlayer coupling leads to a large splitting
of the states at the $\Gamma$ point, with one of them becoming the
valence band maximum. This also occurs for bulk MoS$_{2}$\citep{Cappelluti_2013}
and has been predicted by \textit{ab initio} calculations for commensurate
twisted bilayer structures for several twist angles \citep{Huang_2014,Liu_2014,Yeh_2016}.
We checked, that if the interlayer distance is kept fixed, this splitting
at $\Gamma$ is virtually independent of the twist angle. From this,
it can be inferred that the change in the splitting at $\Gamma$ with
angle predicted in \citep{Huang_2014,Liu_2014,Yeh_2016} and observed
in the ARPES measurements of \citep{Yeh_2016} is due to the interlayer
separation modulation with the twist angle, which we kept fixed. As
we can see in Fig.~\ref{fig:arpes-tBMoS2}(a), the effect of the
interlayer coupling is negligible at the K point. As in the case of
twisted bilayer graphene, the back-folded bands have negligible visibility
in ARPES. In Fig.~\ref{fig:arpes-tBMoS2}(c), we show the constant
energy map for the same $\theta=13.5^{\circ}$ twist angle. For comparison,
the ARPES constant energy map for the decoupled layer is shown in
Fig.~\ref{fig:arpes-tBMoS2}(b). Once again, it can be seen that
the interlayer coupling affects more strongly the states close to
the $\Gamma$ point having almost no impact on the states close to
the K points. It is observed that the valence band pocket at the $\Gamma$
point has very weak visibility in ARPES in agreement with what is
experimentally observed in Ref.~\citep{Yeh_2016}.

\section{Conclusions\label{sec:Conclusions}}

In this work we developed a general theoretical framework to model
ARPES of lattice mismatch/misaligned vdW structures. By describing
the photoemitted state as a plane wave and the bound electronic states
in terms of Bloch waves of the individual layers, while taking into
account generalized umklapp processes, we obtained an efficient description
of ARPES that can be applied both to commensurate and incommensurate
structures. Being based on a tight-binding description of the bound
electronic states, the present formalism can deal with arbitrary lattice
mismatch/misalignment, going beyond previous low energy, continuum
descriptions of both twisted bilayer graphene and graphene/$h$-BN
structures. We applied the developed formalism to the cases of twisted
bilayer graphene and twisted bilayer MoS$_{2}$. The example of graphene
showcases the importance of the ARPES matrix elements in two ways:
(i) by showing the importance of the momentum dependence of the ARPES
weight in entangled bands, which is responsible for the absence of
part of the constant energy surface map around the K points, and (ii)
by showing that the ARPES weight of back folded bands is very small,
which is a consequence of the weak coupling between the two layers.
As a consequence, the observed ARPES bands mostly follow the band
structure of the decoupled system, except at energies at which states
from both layers are degenerate, where a significant reconstruction
of the spectrum occurs. By comparing the results of the current model
to the experimental data of ARPES of twisted bilayer graphene at a
twist angle $\theta=11.6^{\circ}$\citep{Ohta_2012}, we showcased
the importance of the transferred momentum between the incoming radiation
and the photoemitted state along the $z$ direction, which significantly
affects the visibility of bands in ARPES that correspond to bonding
and anti-bonding states of the two layers. In the example of MoS$_{2}$,
we once again observed that the ARPES bands mostly follow the band
structure of the decoupled system. Differently from graphene, we found
out that while there is a significant shift in energy of the states
at the $\Gamma$ point, the reconstruction of the valence band close
to the K points due to the interlayer coupling is negligible, which
is in accordance with experimental ARPES data \citep{Yeh_2016}. 

Although this work focused on bilayer vdW structures, the formalism
can also be applied to structures formed by multiple lattice misaligned/mismatch
layers. We would like to point out that the present approach to model
ARPES can also be extended to other systems where competition between
different periodicities occurs, both commensurate and incommensurate,
such as 2D materials placed on top of a weakly coupled substrate or
materials displaying CDW phases. The coupling to a substrate can be
described via a substrate surface Green's function, which would play
a role similar to $H_{b}$ in the present work. The effect of the
substrate can also be approximated by a potential which acts on the
2D material. In systems with CDW phases, two cases must be distinguished:
(i) a system with weak fluctuations, which can be described at a single-particle
level, and (ii) a strongly interacting system, with strong fluctuations.
In the former case, it is possible to describe the system with a single-particle
(mean-field) Hamiltonian as done in this work. In the latter case,
it is not possible to describe the system at a single-particle level
and the electronic spectral function of crystal bound states, which
can be severely reconstructed by the interactions, has to be obtained
from the full Green's function of the interacting system\citep{Flicker_2016}.
In this case, it is therefore crucial to distinguish between the effects
of interactions and the effect of the ARPES matrix elements (which
encode the effect of competition between the periodicities) \citep{Kordyuk_2014}.
The approach presented in this work can be used to take into account
the latter effect.
\begin{acknowledgments}
The author would like to acknowledge useful discussions with J. L.
Lado, J. Fernández-Rossier and Eduardo V. Castro. The author also
acknowledges the hospitality of the International Iberian Nanotechnology
Laboratory (INL), where early ideas on this problem were developed.
B. A. received funding from the European Union's Horizon 2020 research
and innovation programme under grant agreement No 706538. 
\end{acknowledgments}

\appendix

\section{Non-equilibrium Green's function description of ARPES\label{appx:ARPES_derivation}}

In this appendix we briefly review the rigorous derivation of the
observed photoemitted current in an ARPES experiment using the non-equilibrium
Green's function technique. This approach was first developed in Ref.~\citep{Caroli_1973}
and its connection to other formalisms and experimental observations
clarified in Ref.~\citep{Feibelman_1974}.

In order to describe the process of photoemission, we must consider
both states that are bound to the crystal and unbound, nearly free,
states. Therefore, in the model employed, the crystal must not occupy
the entire space. For concreteness we assume that the positions of
the atoms that form the crystal are restricted to the $z\leq0$ region.
We assume that the crystal is infinite along the $x-y$ plane. The
single particle Hamiltonian for the electrons reads 
\begin{equation}
H_{0}=\int d^{3}\mathbf{r}\Psi^{\dagger}(\mathbf{r})\left[\frac{\mathbf{p}^{2}}{2m}+V(\mathbf{r})\right]\Psi(\mathbf{r}),\label{eq:Hamiltonian_semi_infinite}
\end{equation}
where $\Psi^{\dagger}(\mathbf{r})$ is the electronic creation field
operator, $V(\mathbf{r})$ is the crystal potential, and $\mathbf{p}=-i\hbar\nabla_{\mathbf{r}}$.
We describe the exciting electromagnetic field in the Weyl gauge,
such that $\mathbf{E}(t,\mathbf{r})=-\partial_{t}\mathbf{A}(t,\mathbf{r})$
and $\mathbf{B}(t,\mathbf{r})=\nabla\times\mathbf{A}(t,\mathbf{r})$,
where $\mathbf{A}(t,\mathbf{r})$ is the vector potential. The coupling
to the electromagnetic field is obtained via minimal coupling $\mathbf{p}\rightarrow\mathbf{p}+e\mathbf{A}(t,\mathbf{r})$.
In the presence of the vector potential, the Hamiltonian can be written
as the sum of three terms 
\begin{equation}
H=H_{0}+H_{1,\mathbf{A}}+H_{2,\mathbf{A}},
\end{equation}
where $H_{0}$ is the same as in Eq.~(\ref{eq:Hamiltonian_semi_infinite})
and 
\begin{align}
H_{1,\mathbf{A}} & =-\int d^{3}\mathbf{r}\left(\Psi^{\dagger}(\mathbf{r})\mathbf{J}^{(1)}(\mathbf{r})\Psi(\mathbf{r})\right)\cdot\mathbf{A}(\mathbf{r},t),\\
H_{2,\mathbf{A}} & =-\frac{1}{2}\int d^{3}\mathbf{r}\left(\Psi^{\dagger}(\mathbf{r})J^{(2)}(\mathbf{r})\Psi(\mathbf{r})\right)\mathbf{A}^{2}(\mathbf{r},t),
\end{align}
with 
\begin{align}
\mathbf{J}^{(1)}(\mathbf{r}) & =-\frac{e\hbar}{2mi}\left(\overrightarrow{\nabla}_{\mathbf{r}}-\overleftarrow{\nabla}_{\mathbf{r}}\right),\\
J^{(2)}(\mathbf{r}) & =-\frac{e^{2}}{m},
\end{align}
the matrix elements in real space of the paramagnetic and diamagnetic
currents, respectively. The arrows in the differential operator indicate
whether the derivative acts on the electronic field operator placed
to the right or to the left. The expectation value of the current
is given by 
\begin{multline}
\left\langle \mathbf{J}(t,\mathbf{r})\right\rangle =\lim_{\mathbf{r}^{\prime}\rightarrow\mathbf{r}}\left[-e\frac{\hbar}{2mi}\left(\nabla_{\mathbf{r}}-\nabla_{\mathbf{r}^{\prime}}\right)-\frac{e^{2}}{m}\mathbf{A}(t,\mathbf{r})\right]\times\\
\times\left(-i\right)G_{\mathbf{A}}^{<}(t,\mathbf{r};t,\mathbf{r}^{\prime}),\label{eq:current}
\end{multline}
with the lesser Green's function defined as 
\begin{equation}
G_{\mathbf{A}}^{<}(t,\mathbf{r};t^{\prime},\mathbf{r}^{\prime})=i\left\langle \Psi^{\dagger}(t^{\prime},\mathbf{r}^{\prime})\Psi(t,\mathbf{r})\right\rangle _{\mathbf{A}},
\end{equation}
where the $\mathbf{A}$ subscript means the expectation value is evaluated
taking into account $\mathbf{A}(t,\mathbf{r})$. This can be done
perturbatively in $\mathbf{A}(t,\mathbf{r})$ using the non-equilibrium
Green's function formalism \citep{HaugPekka_book,Rammer_book_neq}.
By expanding the contour-ordered Green's function in the Schwinger-Keldysh
contour in powers of $\mathbf{A}(t,\mathbf{r})$ and then using Langreth's
rules, the lesser Green's function is given to second order in $\mathbf{A}(t,\mathbf{r})$
by 
\begin{align}
\bm{G}_{\mathbf{A}}^{<} & =\bm{G}^{<}+\bm{G}^{<}\odot\left(\mathbf{J}^{(1)}\cdot\mathbf{A}\right)\odot\bm{G}^{A}\nonumber \\
 & +\bm{G}^{R}\odot\left(\mathbf{J}^{(1)}\cdot\mathbf{A}\right)\odot\bm{G}^{<}\nonumber \\
 & +\frac{1}{2}\bm{G}^{<}\odot\left(J^{(2)}\mathbf{A}^{2}\right)\odot\bm{G}^{A}+\nonumber \\
 & +\frac{1}{2}\bm{G}^{R}\odot\left(J^{(2)}\mathbf{A}^{2}\right)\odot\bm{G}^{<}\nonumber \\
 & +\bm{G}^{<}\odot\left(\mathbf{J}^{(1)}\cdot\mathbf{A}\right)\cdot\bm{G}^{A}\cdot\left(\mathbf{J}^{(1)}\cdot\mathbf{A}\right)\odot\bm{G}^{A}\nonumber \\
 & +\bm{G}^{R}\odot\left(\mathbf{J}^{(1)}\cdot\mathbf{A}\right)\odot\bm{G}^{R}\odot\left(\mathbf{J}^{(1)}\cdot\mathbf{A}\right)\odot\bm{G}^{<}\nonumber \\
 & +\bm{G}^{R}\odot\left(\mathbf{J}^{(1)}\cdot\mathbf{A}\right)\odot\bm{G}^{<}\odot\left(\mathbf{J}^{(1)}\cdot\mathbf{A}\right)\odot\bm{G}^{A}\nonumber \\
 & +\mathcal{O}\left(\mathbf{A}^{3}\right)\label{eq:lesser_G_2nd_order_expansion}
\end{align}
where the retarded and advanced Green's functions are given by 
\begin{align}
G^{R}\left(t,\mathbf{r};t^{\prime},\mathbf{r}^{\prime}\right) & =-i\Theta(t-t^{\prime})\left\langle \left\{ \Psi(t,\mathbf{r}),\Psi^{\dagger}(t^{\prime},\mathbf{r}^{\prime})\right\} \right\rangle ,\\
G^{A}\left(t,\mathbf{r};t^{\prime},\mathbf{r}^{\prime}\right) & =i\Theta(t^{\prime}-t)\left\langle \left\{ \Psi(t,\mathbf{r}),\Psi^{\dagger}(t^{\prime},\mathbf{r}^{\prime})\right\} \right\rangle ,
\end{align}
and the $\odot$ product represents integration over the spatial variables
and time, and summation over other possible degrees of freedom (sublattice,
orbital, spin,...). All the Green's functions in Eq.~(\ref{eq:lesser_G_2nd_order_expansion})
are evaluated in the absence of $\mathbf{A}(t,\mathbf{r})$, and are
therefore in thermodynamic equilibrium. In Eq.~(\ref{eq:lesser_G_2nd_order_expansion})
interactions between the emitted state and the remaining hole are
neglected, an approximation that is typically referred to as the sudden
approximation. Including these kind of interactions would lead to
a renormalization of the vertices $\mathbf{J}^{(1)}$ and $J^{(2)}$
\citep{Caroli_1973}. For a photodectector placed very far away from
the crystal sample, only the last term in Eq.~(\ref{eq:lesser_G_2nd_order_expansion})
gives a finite contribution. This can be understood if we write $G^{<}$
in terms of the eigenstates $\psi_{a}(\mathbf{r})$ of Eq.~(\ref{eq:Hamiltonian_semi_infinite}).
The creation field operator can be written as $\Psi^{\dagger}(\mathbf{r})=\sum_{a}c_{a}^{\dagger}\psi_{a}^{*}(\mathbf{r})$,
where $c_{a}^{\dagger}$ creates an electron in state $\psi_{a}(\mathbf{r})$
with energy $\epsilon_{a}$, and the lesser Green's function, for
a non-interacting system, becomes 
\begin{equation}
G^{<}(t,\mathbf{r};t^{\prime},\mathbf{r}^{\prime})=\sum_{a}\psi_{a}(\mathbf{r})\psi_{a}^{*}(\mathbf{r}^{\prime})e^{-i\epsilon_{a}\left(t-t^{\prime}\right)}i\left\langle c_{a}^{\dagger}c_{a}\right\rangle .
\end{equation}
Notice that only crystal bound states are occupied, and therefore
the sum in previous equations is restricted to those states due to
the occupation function $\left\langle c_{a}^{\dagger}c_{a}\right\rangle $.
At the same time, the wavefunction of crystal bound states decays
exponentially away from the crystal. Therefore if one of the arguments
of $G^{<}$ is evaluated away from the crystal, that term can be safely
neglected. This also allows us to discard the diamagnetic term in
Eq.~(\ref{eq:current}), as it would only contribute to third order
in $\mathbf{A}(t,\mathbf{r})$. Therefore, to second order in the
electromagnetic field, the photoemitted current measured away from
the crystal is given by

\begin{multline}
\left\langle \mathbf{J}(t,\mathbf{r})\right\rangle =\lim_{\mathbf{r}^{\prime}\rightarrow\mathbf{r}}\frac{e\hbar}{2m}\left(\nabla_{\mathbf{r}}-\nabla_{\mathbf{r}^{\prime}}\right)\int dt_{1}d^{3}\mathbf{r}_{1}\int dt_{2}d^{3}\mathbf{r}_{2}\\
G^{R}(t,\mathbf{r};t_{1},\mathbf{r}_{1})\mathbf{J}^{(1)}(\mathbf{r}_{1})\cdot\mathbf{A}(t_{1},\mathbf{r}_{1})\times\\
\times G^{<}(t_{1},\mathbf{r}_{1};t_{2},\mathbf{r}_{2})\mathbf{J}^{(1)}(\mathbf{r}_{2})\cdot\mathbf{A}(t_{2},\mathbf{r}_{2})G^{A}(t_{2},\mathbf{r}_{2};t,\mathbf{r}^{\prime}).
\end{multline}
Assuming a monochromatic electromagnetic field at frequency $\omega_{0}>0$,
\begin{equation}
\mathbf{A}(t,\mathbf{r})=\mathbf{A}_{\omega_{0}}(\mathbf{r})e^{-i\omega_{0}t}+\mathbf{A}_{-\omega_{0}}(\mathbf{r})e^{i\omega_{0}t},
\end{equation}
expressing all quantities in Fourier components in time and looking
at the time-averaged current over one period $2\pi/\omega_{0}$, we
obtain 
\begin{multline}
\overline{\left\langle \mathbf{J}(\mathbf{r})\right\rangle }=\lim_{\mathbf{r}^{\prime}\rightarrow\mathbf{r}}\frac{e\hbar}{2m}\left(\nabla_{\mathbf{r}}-\nabla_{\mathbf{r}^{\prime}}\right)\int\frac{dE}{2\pi}\int d^{3}\mathbf{r}_{1}\int d^{3}\mathbf{r}_{2}\\
\left[G^{R}(E;\mathbf{r},\mathbf{r}_{1})\mathbf{J}^{(1)}(\mathbf{r}_{1})\cdot\mathbf{A}_{\omega_{0}}(\mathbf{r}_{1})G^{<}\left(E-\omega_{0};\mathbf{r}_{1},\mathbf{r}_{2}\right)\right.\times\\
\left.\times\mathbf{J}^{(1)}(\mathbf{r}_{2})\cdot\mathbf{A}_{-\omega_{0}}(\mathbf{r}_{2})G^{A}(E;\mathbf{r}_{2},\mathbf{r}^{\prime})+\left(\omega_{0}\leftrightarrow-\omega_{0}\right)\right].\label{eq:averaged_current}
\end{multline}
Some further simplifications can be performed. First, by using the
fluctuation-dissipation theorem, which is valid in thermodynamic equilibrium
even for an interacting system, we can write 
\begin{equation}
G^{<}\left(E;\mathbf{r}_{1},\mathbf{r}_{2}\right)=\sum_{a}\psi_{a}(\mathbf{r}_{1})\psi_{a}^{*}(\mathbf{r}_{2})if\left(E-\mu\right)\mathcal{A}_{a}\left(E\right),\label{eq:fluctuation_dissipation}
\end{equation}
where $\mathcal{A}_{a}\left(E\right)=G_{a}^{R}(E)-G_{a}^{A}(E)$ is
the electronic spectral function in the eigenstate basis, and $f(\omega)=\left(e^{\beta\omega}+1\right)^{-1}$
is the Fermi-Dirac distribution function with $\mu$ the chemical
potential. Once again, only crystal bound states are occupied and
therefore, the integration over the spatial coordinates in Eq.~\eqref{eq:averaged_current}
is mostly confined to the region of the crystal. This allows us to
use the asymptotic expression of the retarded and advanced Green's
functions, valid for $\mathbf{r}$ and $\mathbf{r}^{\prime}$ far
away from the crystal, \citep{Adawi_1964,Feibelman_1974} 
\begin{align}
G^{R}(E;\mathbf{r},\mathbf{r}_{1}) & \simeq-\frac{2m}{\hbar^{2}}\frac{1}{4\pi r}e^{ip_{E}r}\psi_{E,\hat{\mathbf{r}}}^{*}(\mathbf{r}_{1}),\label{eq:retarded_GF_far_field}\\
G^{A}(E;\mathbf{r}_{2},\mathbf{r}^{\prime}) & \simeq-\frac{2m}{\hbar^{2}}\frac{1}{4\pi r^{\prime}}\psi_{E,\hat{\mathbf{r}}}\left(\mathbf{r}_{2}\right)e^{-ip_{E}^{*}r^{\prime}},\label{eq:advanced_GF_far_field}
\end{align}
where $p_{E}=\sqrt{2mE/\hbar^{2}}$ for $E>0$ and $p_{E}=i\sqrt{2m\Bigl|E\Bigr|/\hbar^{2}}$
for $E<0$ (we choose as zero of energy the threshold to have free
electron states), $\psi_{E,\hat{\mathbf{r}}}^{*}(\mathbf{r}_{1})$,
given by Eq.~(\ref{eq:inverse_difraction_state}), is the conjugate
of an electron diffraction state \citep{Adawi_1964,Feibelman_1974},
$\hat{\mathbf{r}}$ is the unit vector pointing along $\mathbf{r}$,
and $G_{\text{free}}^{R}(E;\mathbf{r},\mathbf{r}^{\prime})$ is the
free space retarded Green's function 
\begin{equation}
G_{\text{free}}^{R}(E;\mathbf{r},\mathbf{r}^{\prime})=-\frac{2m}{\hbar^{2}}\frac{1}{4\pi}\frac{e^{ip_{E}\left|\mathbf{r}-\mathbf{r}^{\prime}\right|}}{\left|\mathbf{r}-\mathbf{r}^{\prime}\right|}.\label{eq:free_space_GF}
\end{equation}
Equation (\ref{eq:retarded_GF_far_field}) can be obtained from the
Dyson equation for the retarded Green's function 
\begin{multline}
G^{R}(E;\mathbf{r},\mathbf{r}^{\prime})=G_{\text{free}}^{R}(E;\mathbf{r},\mathbf{r}^{\prime})+\\
+\int d^{3}\mathbf{r}_{1}G_{\text{free}}^{R}(E;\mathbf{r},\mathbf{r}_{1})V(\mathbf{r}_{1})G^{R}(E;\mathbf{r}_{1},\mathbf{r}^{\prime}).\label{eq:Dyson_GF}
\end{multline}
If we are interested in the Green's function when $\mathbf{r}$ is
very far away from the crystal, we can use the following approximation
for $G_{\text{free}}^{R}(E;\mathbf{r},\mathbf{r}^{\prime})$, valid
for $\left|\mathbf{r}\right|\gg\left|\mathbf{r}^{\prime}\right|$,
\begin{equation}
G_{\text{free}}^{R}(E;\mathbf{r},\mathbf{r}^{\prime})\simeq-\frac{2m}{\hbar^{2}}\frac{1}{4\pi}\frac{e^{ip_{E}r}}{r}e^{-ip_{R}\hat{\mathbf{r}}\cdot\mathbf{r}^{\prime}}.
\end{equation}
Inserting this approximation into the Dyson equation (\ref{eq:Dyson_GF}),
we obtain Eq.~(\ref{eq:retarded_GF_far_field}) with 
\begin{equation}
\psi_{E,\hat{\mathbf{r}}}^{*}(\mathbf{r}^{\prime})=e^{-ip_{E}\hat{\mathbf{r}}\cdot\mathbf{r}^{\prime}}+\int d^{3}\mathbf{r}_{1}e^{-ip_{E}\hat{\mathbf{r}}\cdot\mathbf{r}_{1}}V(\mathbf{r}_{1})G^{R}(E;\mathbf{r}_{1},\mathbf{r}^{\prime}).
\end{equation}
Using the Dyson equation for the Green's function Eq.~(\ref{eq:Dyson_GF}),
we can rewrite the above equation as Eq.~(\ref{eq:inverse_difraction_state}),
which is nothing more than the Lippmann-Schwinger equation for the
scattering of an incoming plane wave state, $\phi_{0}^{*}(\mathbf{r}^{\prime})=e^{-ip_{E}\hat{\mathbf{r}}\cdot\mathbf{r}^{\prime}}$,
by the crystal potential $V(\mathbf{r})$. Equation (\ref{eq:advanced_GF_far_field})
for the advanced Green's function can be obtained in a similar way.
Inserting Eqs.~(\ref{eq:fluctuation_dissipation}), (\ref{eq:retarded_GF_far_field})
and (\ref{eq:advanced_GF_far_field}) in Eq.~(\ref{eq:averaged_current}),
the detected photoemitted current becomes 
\begin{multline}
\overline{\left\langle \mathbf{J}(\mathbf{r})\right\rangle }=\frac{-e}{\left(4\pi r\right)^{2}}\hat{\mathbf{r}}\int\frac{dE}{2\pi}\text{Re}\left[\frac{\hbar p_{E}}{m}e^{i\left(p_{E}-p_{E}^{*}\right)r}\right]\times\\
\times\sum_{a}\left[f\left(E-\omega_{0}-\mu\right)\left|M_{E,\hat{\mathbf{r}};a}(\omega_{0})\right|^{2}\mathcal{A}_{a}\left(E-\omega_{0}\right)+\right.\\
\left.+\left(\omega_{0}\leftrightarrow-\omega_{0}\right)\right].\label{eq:ARPES_current_beta}
\end{multline}
where we have defined the ARPES matrix elements as 
\begin{multline}
M_{E,\mathbf{n};a}(\omega_{0})=-\frac{2m}{\hbar^{2}}\times\\
\times\int d^{3}\mathbf{r}_{1}\left(\psi_{E,\mathbf{n}}^{*}(\mathbf{r}_{1})\mathbf{J}^{(1)}(\mathbf{r}_{1})\psi_{a}(\mathbf{r}_{1})\right)\cdot\mathbf{A}_{\omega_{0}}(\mathbf{r}_{1}),\label{eq:ARPES_matrix_elements_appendix}
\end{multline}
which can also be written as Eq.~(\ref{eq:ARPES_matrix_elements})
of the main text. For $E<0$, we have that $e^{i\left(p_{E}-p_{E}^{*}\right)r}=e^{-2\left|p_{E}\right|r}$
and therefore this contribution vanishes for a detector far away from
the crystal. Therefore, we can restrict the integration in Eq.~(\ref{eq:ARPES_current_beta})
from $0$ to $+\infty$. At the same time, we notice that the second
term in Eq.~(\ref{eq:ARPES_current_beta}) involves states with energy
$E+\omega_{0}>0$ which are unoccupied, and can therefore be neglected.
This allows us to write 
\begin{multline}
\overline{\left\langle \mathbf{J}(\mathbf{r})\right\rangle }=\frac{-e}{\left(4\pi r\right)^{2}}\hat{\mathbf{r}}\int_{0}^{+\infty}\frac{dE}{2\pi}\frac{\hbar p_{E}}{m}f\left(E-\omega_{0}-\mu\right)\times\\
\times\sum_{a}\left|M_{E,\hat{\mathbf{r}};a}(\omega_{0})\right|^{2}\mathcal{A}_{a}\left(E-\omega_{0}\right).
\end{multline}
Assuming that the electron detector can resolve the energy of the
photoemitted states and it collects electrons emitted along direction
$\hat{\mathbf{r}}$, the measured ARPES intensity is proportional
to Eq.~(\ref{eq:ARPES_intensity}) of the main text.

\section{Fourier transform of interlayer hopping for MoS$_{2}$\label{appx:interlayer_coupling}}

In this appendix, we provide details on how the two-dimensional Fourier
transform of the interlayer coupling for S $p$ orbitals for twisted
bilayer MoS$_{2}$, Eq.~\eqref{eq:pp_interlayer}, is computed. We
wish to evaluate
\begin{multline}
h_{p_{i}^{\text{bot}},p_{j}^{\text{top}}}^{tb}(\mathbf{q})=\int\frac{d^{2}\mathbf{r}}{A_{\text{c}}}e^{-i\mathbf{q}\cdot\mathbf{r}}\times\\
\left[\begin{array}{ccc}
v_{1}(r)+v_{2}(r)\frac{x^{2}}{R^{2}} & v_{2}(r)\frac{xy}{R^{2}} & v_{2}(r)\frac{xd}{R^{2}}\\
v_{2}(r)\frac{xy}{R^{2}} & v_{1}(r)+v_{2}(r)\frac{y^{2}}{R^{2}} & v_{2}(r)\frac{yd}{R^{2}}\\
v_{2}(r)\frac{xd}{R^{2}} & v_{2}(r)\frac{yd}{R^{2}} & v_{1}(r)+v_{2}(r)\frac{d^{2}}{R^{2}}
\end{array}\right].\label{eq:FT}
\end{multline}
where $v_{1}(r)=V_{pp\pi}\left(R\right)$ and $v_{2}(r)=V_{pp\sigma}\left(R\right)-V_{pp\pi}\left(R\right)$,
with $R=\sqrt{r^{2}+d^{2}}$, and $A_{\text{c}}$ is the unit cell
area of MoS$_{2}$. We have three kinds of integrals:
\begin{align}
\mathcal{I}_{0}(\mathbf{q}) & =\int d^{2}\mathbf{r}e^{-i\mathbf{q}\cdot\mathbf{r}}f(r),\\
\mathcal{I}_{i}(\mathbf{q}) & =\int d^{2}\mathbf{r}e^{-i\mathbf{q}\cdot\mathbf{r}}x_{i}f(r),\\
\mathcal{I}_{ij}(\mathbf{q}) & =\int d^{2}\mathbf{r}e^{-i\mathbf{q}\cdot\mathbf{r}}x_{i}x_{j}f(r),
\end{align}
where $f(r)$ is a function only of $r=\left|\mathbf{r}\right|$.
Using the fact that $\int_{0}^{2\pi}d\phi e^{-ix\cos\phi}=2\pi J_{0}(x)$,
where $J_{0}(x)$ is the Bessel function of first kind and order $0$,
we can write
\begin{equation}
\mathcal{I}_{0}(\mathbf{q})=2\pi\int drrf(r)J_{0}(qr).
\end{equation}
The integrals $\mathcal{I}_{i}(\mathbf{q})$ and $\mathcal{I}_{ij}(\mathbf{q})$
can be written as
\begin{align}
\mathcal{I}_{i}(\mathbf{q}) & =i\int d^{2}\mathbf{r}\left(\frac{\partial}{\partial q_{i}}e^{-i\mathbf{q}\cdot\mathbf{r}}\right)f(r)\nonumber \\
 & =i\int drrf(r)\frac{\partial}{\partial q_{i}}J_{0}(qr),\\
\mathcal{I}_{ij}(\mathbf{q}) & =-\int d^{2}\mathbf{r}\left(\frac{\partial^{2}}{\partial q_{i}\partial q_{j}}e^{-i\mathbf{q}\cdot\mathbf{r}}\right)f(r)\nonumber \\
 & =-\int drrf(r)\frac{\partial^{2}}{\partial q_{i}\partial q_{j}}J_{0}(qr).
\end{align}
Using the properties of Bessel functions, it is possible to write
\begin{align}
\frac{\partial}{\partial q_{i}}J_{0}(qr) & =-\frac{1}{2}r^{2}q_{i}\left(J_{0}(qr)+J_{2}(qr)\right),\\
\frac{\partial^{2}}{\partial q_{i}\partial q_{j}}J_{0}(qr) & =-\delta_{ij}\frac{1}{2}r^{2}\left(J_{0}(qr)+J_{2}(qr)\right)\nonumber \\
 & +\frac{q_{i}q_{j}}{q^{2}}r^{2}J{}_{2}(qr).
\end{align}
Using this, we can write Eq.~\eqref{eq:FT} as Eq.~\eqref{eq:interlayer_MoS2},
with
\begin{align}
V_{1}(q) & =\frac{2\pi}{A_{\text{c}}}\int drr\biggl[v_{1}(r)J_{0}(qr),\nonumber \\
 & \left.+\frac{1}{2}v_{2}(r)\frac{r^{2}}{R^{2}}\left(J_{0}(qr)+J_{2}(qr)\right)\right]\\
V_{2}(q) & =\frac{2\pi}{A_{\text{c}}}\int drrv_{2}(r)\frac{r^{2}}{R^{2}}J_{2}(qr),\\
V_{z}(q) & =qd\frac{\pi}{A_{\text{c}}}\int drrv_{2}(r)\frac{r^{2}}{R^{2}}\left[J_{0}(qr)+J_{2}(qr)\right],\\
V_{zz}(q) & =\frac{2\pi}{A_{\text{c}}}\int drr\left[v_{1}(r)+v_{2}(r)\frac{d^{2}}{R^{2}}\right]J_{0}(qr).
\end{align}
The remaining integration over $r$ has to be performed numerically. 

%

\end{document}